\documentclass[longauth]{aa} 
\usepackage{graphicx}
\usepackage{color}
\usepackage{longtable}
\usepackage{supertabular}
\usepackage{natbib}

\definecolor{linkcolor}{rgb}{0.2, 0.55, 0.9}
\usepackage[breaklinks, colorlinks, citecolor=linkcolor]{hyperref}

\usepackage[varg]{txfonts}

\defcitealias{MBB13}{MBB} 
\defcitealias{Biviano+16}{B+16}
\begin{document}

\definecolor{red}{rgb}{1.,0.,0.}
\definecolor{nnew}{rgb}{0.6,0.2,0.0}
\newcommand{\rvdb}[1]{{\color{rvdb} #1}}
\newcommand{\nnew}[1]{\textcolor{nnew}{\bf #1}}

\newcommand{\mr}{M(r)}
\newcommand{\br}{\beta(r)}
\newcommand{\bpr}{\sigma_{\rm r}/\sigma_{\theta}}
\newcommand{\rvir}{r_{200}}
\newcommand{\rtwo}{r_{-2}}
\newcommand{\cvir}{c_{200}}
\newcommand{\vvir}{v_{200}}
\newcommand{\mvir}{M_{200}}
\newcommand{\rnu}{r_{\nu}}
\newcommand{\ks}{\mathrm{km \, s}^{-1}}
\newcommand{\slos}{\sigma_{\rm{los}}}
\newcommand{\vrf}{v_{{\rm rf}}}
\newcommand{\qr}{Q_{{\rm r}}}
\newcommand{\qt}{Q_{{\rm th}}}
\newcommand{\qtr}{Q_{{\rm r,th}}}
\newcommand{\ar}{\alpha_{\rm{r}}}
\newcommand{\ab}[1]{\textcolor{magenta}{\bf ABi: #1}}

\title{The GOGREEN survey: Internal dynamics of clusters of galaxies at redshift $0.9-1.4$}
\author{
  A. Biviano \inst{1,2}, R. F. J. van der Burg \inst{3}, Michael L. Balogh \inst{4,5}, E. Munari \inst{1}, 
  M. C. Cooper\inst{6}, G. De Lucia\inst{1},  R. Demarco \inst{7}, P. Jablonka \inst{8}, A. Muzzin \inst{9}, J. Nantais\inst{10}, L. J. Old\inst{11}, G. Rudnick \inst{12}, B. Vulcani \inst{13},  G. Wilson \inst{14}, H. K. C. Yee\inst{15}, D. Zaritsky \inst{16}, P. Cerulo \inst{23}, J. Chan \inst{14}, A. Finoguenov\inst{17}, D. Gilbank \inst{18,19}, C. Lidman \inst{20,21}, I. Pintos-Castro \inst{15}, H. Shipley \inst{22}
}
\offprints{Andrea Biviano, andrea.biviano@inaf.it}
          
\institute{
  INAF-Osservatorio Astronomico di Trieste, via G. B. Tiepolo 11, 
  I-34131, Trieste, Italy
\and 
             IFPU-Institute for Fundamental Physics of the Universe, via Beirut 2, 34014 Trieste, Italy
 \and
 European Southern Observatory, Karl-Schwarzschild-Str. 2, 85748, Garching, Germany
\and 
Department of Physics and Astronomy, University of Waterloo, Waterloo, Ontario, N2L 3G1, Canada
\and
Waterloo Centre for Astrophysics, University of Waterloo, Waterloo, Ontario, N2L 3G1, Canada
\and
Department of Physics \& Astronomy, University of California Irvine, 4129 Reines Hall, Irvine, CA 92697, USA
 \and
 Departamento de Astronom\'ia, Facultad de Ciencias F\'isicas y Matem\'aticas,
Universidad de Concepci\'on, Concepci\'on, Chile 
 \and 
 Physics Institute, Laboratory of Astrophysics, Ecole Polytechnique F\'ed\'erale de Lausanne (EPFL), 1290 Sauverny, Switzerland
 \and
 Department of Physics and Astronomy, York University, 4700 Keele Street, Toronto, Ontario, ON MJ3 1P3, Canada
\and
Departamento de Ciencias F\'isicas, Universidad Andres Bello, Fern\'andez Concha 700, Las Condes 7591538, RM, Chile
\and
European Space Agency (ESA), European Space Astronomy Centre, Villanueva de la Ca\~{n}ada, E-28691 Madrid, Spain
\and
Department of Physics and Astronomy, The University of Kansas, Malott room 1082, 1251 Wescoe Hall Drive, Lawrence, KS 66045, USA
 \and
 INAF- Osservatorio astronomico di Padova, Vicolo Osservatorio 5, 35122 Padova, Italy
\and
Department of Physics and Astronomy, University of California Riverside, 900 University Avenue, Riverside, CA 92521, USA
\and
Department of Astronomy and Astrophysics, University of Toronto, Toronto, ON M5S 2H4, Canada
\and
Steward Observatory, University of Arizona, 933 N. Cherry Ave., Tucson, AZ, USA
\and
Department of Physics, Gustaf Hällströmin katu 2 A, University of Helsinki, Helsinki, Finland
\and
South African Astronomical Observatory, PO Box 9, Observatory, 7935, South Africa
\and
Centre for Space Research, North-West University, Potchefstroom 2520, South Africa\\
\and
Research School of Astronomy and Astrophysics, The Australian National University, ACT 2601, Australia
\and
Centre for Gravitational Astrophysics, College of Science, The Australian National University, ACT 2601, Australia
\and
Department of Physics, McGill University, 3600 rue University, Montr\'{e}al, Qu\'{e}bec, H3P 1T3, Canada
\and Departamento de Ingenier\'ia Inform\'atica y Ciencias de la Computaci\'on, Facultad de Ingenier\'ia, Universidad de Concepci\'on, Concepci\'on, Chile
}
\date{}

\abstract
{The study of galaxy cluster mass profiles ($M(r)$) provides constraints on the nature of dark matter and on physical processes affecting the mass distribution. The study of galaxy cluster velocity anisotropy profiles ($\beta(r)$) informs the orbits of galaxies in clusters, which are related to their evolution. The combination of mass profiles and velocity anisotropy profiles allows us to determine the pseudo phase-space density profiles ($Q(r)$);  numerical simulations predict that these profiles  follow a simple power law in cluster-centric distance.}
{We determine the mass, velocity anisotropy, and pseudo phase-space density profiles of clusters of galaxies at the highest redshifts investigated in detail to date.}
{We exploited the combination of the GOGREEN and GCLASS spectroscopic data-sets
for 14 clusters with mass $\mvir \geq 10^{14} \, {\rm M}_{\odot}$
at redshifts $0.9 \leq z \leq 1.4$. We constructed an {\em ensemble} cluster by stacking 581 spectroscopically identified cluster members with stellar mass $M_{\star} \geq 10^{9.5} {\rm M}_{\odot}$. We used the MAMPOSSt method to constrain several $M(r)$ and $\beta(r)$ models, and we then inverted the Jeans equation to determine the {\em ensemble} cluster $\beta(r)$ in a non-parametric way. Finally, we combined the results of the $M(r)$ and $\beta(r)$ analysis to determine $Q(r)$ for the {\em ensemble} cluster.}
{The concentration $\cvir$ of the {\em ensemble} cluster mass profile is in excellent agreement with predictions from Lambda cold dark matter ($\Lambda$CDM)  cosmological numerical simulations, and with previous determinations for clusters of similar mass and at similar redshifts, obtained from gravitational lensing and X-ray data. We see no significant difference between the total mass density and either the galaxy number density distributions or the stellar mass distribution. Star-forming galaxies are spatially significantly less concentrated than quiescent galaxies.
The orbits of cluster galaxies are isotropic near the center and more radial outside. Star-forming galaxies and galaxies of low stellar mass tend to move on more radially elongated orbits than quiescent galaxies and galaxies of high stellar mass. The profile $Q(r)$, determined  using either the total mass or the number density profile, is very close to the power-law behavior predicted by numerical simulations. }
{The internal dynamics of clusters at the highest redshift probed in detail to date are very similar to those of lower-redshift clusters, and in excellent agreement with predictions of numerical simulations. The clusters in our sample have already reached a high degree of dynamical relaxation.}


\titlerunning{GOGREEN cluster members}
\authorrunning{A. Biviano}

\maketitle

\section{Introduction}
Cosmological halos from  Lambda cold dark matter ($\Lambda$CDM)
numerical simulations are known to have an internal mass distribution described by a universal shape, well represented by the Navarro-Frenk-White (NFW) model \citep{NFW96,NFW97}. Alternative models of this universal profile have been proposed \citep[e.g.,][]{Moore+98,RPV07,DelPopolo10,Navarro+04,Stadel+09}. The inner slope of the mass distribution might deviate from the NFW model because of several processes such as dynamical friction, central condensation of cooled gas, and active galactic nucleus (AGN)  feedback \citep[e.g.,][]{Blumenthal+86,MTMW12,Schaller+15,LW15,Peirani+17,He+20}. The external slope of the mass distribution in individual clusters might deviate from the NFW model depending on the mass accretion rate in the cluster outskirts \citep[e.g.,][]{DK14,MDK15,VPPQ20}. The shape of the cluster mass profile also depends on the properties of the dark matter (DM) component. Models that deviate from the classical cold DM (CDM) model  generally predict mass density profiles, $\rho(r)$, that are flatter than the NFW model near the halo center \citep[see, e.g.,][]{HBG00,BOT01,PMK10,SS00,Rocha+13}. An example of these profiles is the cored profile of \citet{Burkert95}.

The two quantities that characterize most models for the mass profile of cosmological halos, $M(r),$ are a scale parameter, the mass at a given overdensity, $M_{\Delta}$\footnote{\label{foot:mvir} The mass $M_{\Delta}$ is the mass contained within a sphere of radius $r_{\Delta}$ within which the mean mass overdensity is $\Delta$ times the critical density at the cluster's redshift. In this paper we adopt $\Delta=200$, so we have
$\mvir \equiv 200 \, H_z^2 \, \rvir^3/(2 \, G)$, where $H_z$ is the Hubble parameter at the cluster's redshift $z$. The circular velocity at the radius $\rvir$ is $\vvir=10 \, H_z \, \rvir$.}, and a shape parameter, the concentration $c_{\Delta} \equiv r_{\Delta}/r_{-2}$, with $r_{-2}$  the radius where the logarithmic derivative of the mass density profile $\rho(r)$, $\gamma   \equiv {\rm d}\ln \rho / {\rm d}\ln r =-2$. A prediction of CDM cosmological simulations is the existence of a relation between $c_{\Delta}$ and $M_{\Delta}$ \citep[see, e.g.,][]{NFW96,Bullock+01} that depends on the halo's redshift, $z$. More massive halos are less concentrated because they form later 
when the density of the Universe is lower \citep{WR78}. The concentration is predicted to decrease with increasing $z$ at a given $M_{\Delta}$, and the concentration--mass relation is predicted to flatten with $z$ \citep[e.g.,][]{NFW96,Bullock+01,ZJMB03,Neto+07}. The earliest investigations predicted a strong $z$ dependence of the concentration--mass relation, but more recent works indicate a very mild $z$ dependence, with $c_{\Delta}$ increasing by only $\sim 30$\% from $z \sim 2$ to $z \sim 0$ \citep[e.g.,][]{DeBoni+13,DM14}. The normalization and evolution of the concentration--mass relation depends on the cosmological model, in particular on the Hubble and density parameters $h$ and $\Omega_m$; the dispersion of the mass fluctuation within spheres of comoving radius equal to 8 $h^{-1}$ Mpc, $\sigma_8$; and the dark energy equation of state parameter $w$ \citep[e.g.,][]{KMMB03,Dolag+04,MDvdB08,Carlesi+12,DeBoni+13,KBHH13}. The concentration--mass  relation also depends on baryonic processes. Hydrodynamical simulations have shown that, at a given mass, gas cooling and star formation tend to increase the halo concentration  compared to the case where only gravitational processes are considered \citep{Fedeli12,MTMW12,Rasia+13,Cui+16}. However, feedback from AGN has the opposite effect, so that the concentrations of cosmological halos turn out to be similar in DM-only and full hydrodynamical simulations \citep{KM11,TMMDM11,MTMW12,MTM13,Rasia+13,Schaller+15,SLN18}.

The universal shape of the mass profiles of cosmological halos might be the result of  an initial fast assembly phase \citep[e.g.,][]{HJS99_MN,ElZant08,LC11}, characterized by the chaotic mixing and violent relaxation processes \citep{Henon64,LyndenBell67}. \citet{TN01} suggested that an even more universal quantity is  the pseudo-phase-space density profile, $Q(r) \equiv \rho/\sigma^3$, where $\sigma(r)$ is the total velocity dispersion profile of DM particles. Numerical simulations indicate that $Q(r)$, for cosmological halos with a wide range of masses, follows a power-law behavior with a universal slope \citep[e.g.,][]{TN01,DML05,KKH08}. An alternative formulation of the pseudo-phase density profile is given by $\qr(r) \equiv  \rho/\sigma_{\rm r}^3$ \citep{DML05},
where $\sigma_{\rm r}$ is the radial component of $\sigma$. The $\qr$ profile also follows a power law with radius. The lack of a particular scale radius suggests that $Q(r)$ and $\qr$  have a purely gravitational nature, and are more fundamental than $M(r)$ in describing the internal dynamics of cosmological halos. The profiles $Q(r)$ and $\qr(r)$ have been identified with a sort of gravitational entropy, $K \equiv \sigma^2/\rho^{2/3} = Q^{-2/3}$ \citep[e.g.,][]{LC09}, for its formal analogy with the commonly adopted definition of entropy of the hot intra-cluster medium (ICM), $K_{\rm{ICM}} \equiv k_B \, T/ n_e^{2/3}$ \citep[e.g.,][]{Biffi+17}. Very little evolution is predicted for $Q(r)$, with its power-law slope steepening by $\lesssim 15$\% from $z \sim 0$ to $z \sim 2$ \citep{LC09}.

A relation between the power-law behavior of $Q(r)$ and the NFW  shape of $M(r)$ \citep{NFW96,NFW97} can be established through the Jeans equation of dynamical equilibrium \citep{BT87} if the logarithmic slope of $\rho(r)$ has a linear relation with the velocity anisotropy profile
\begin{equation}
\beta(r) \equiv 1-(\sigma_{\theta}^2+\sigma_{\phi}^2)/(2 \, \sigma_r^2),
\label{e:beta}
\end{equation}
where $\sigma_{\theta}$ and $\sigma_{\phi}$ are the two tangential components of the velocity dispersion, usually assumed to be identical in clusters of galaxies. The existence of such a linear relation was  suggested by \citet{HM06} and interpreted by \citet{Hansen09} in terms of the relative shapes of the radial and tangential velocity distribution functions of bound particles in a halo. Studying the velocity anisotropy profile of cosmological halos  is thus important for an understanding of their internal dynamics.

Constraining $M(r), Q(r),$ and $\beta(r)$ and their evolution can in principle provide useful information on how and when cosmological halos reach dynamical equilibrium, which physical processes are involved, and  the nature of DM. Clusters of galaxies are observational targets of particular interest in this sense because they are expected to be the last halos to achieve virial equilibrium in the CDM scenario \citep{WR78}, so we can hope to trace their evolution close to their formation epoch by observing them at relatively low $z$. The transition epoch between the early fast accretion phase and the late slow accretion phase is indeed predicted to occur at $z \lesssim 1$ for the most massive halos \citep{LC09}. 

Another advantage of studying clusters of galaxies is that their $M(r)$ can be determined over a wide range of scales in several ways, that is, via X-ray observations of the hot ICM, via the Sunyaev-Zeldovich effect \citep{SZ70b}, by the gravitational lensing of galaxies in the cluster background, and by analyzing the distribution of cluster members in projected phase-space.
\citep[see][for a review of these methods]{Pratt+19}. The last method  determines $M(r)$ from the central $\sim50$ kpc to very large radii  \citep[e.g.,][]{RD06,Biviano+13}, and  determines the $\beta(r)$ of cluster galaxies \citep[e.g.,][]{BK04}. Knowledge of
$\beta(r)$ allows us to  determine $\sigma(r)$ and $\sigma_{\rm{r}}$ from the line-of-sight velocity dispersion profile of cluster galaxies, $\slos$. If $\rho(r)$ is known from any of the methods mentioned above,  $Q(r)$ and $\qr(r)$ of the DM components can then be derived, modulo an assumption about the similarity of the galaxies and DM $\slos$ since the DM $\slos$ (and $\sigma$ and $\sigma_{{\rm r}}$) are not observables.

There have been many determinations of the $M(r)$ of clusters of galaxies using cluster galaxies as tracers of the potential, both for individual clusters \citep[e.g.,][]{GDK99,Rines+00,LM03,RGKD03,Lokas+06,RD06,WL10,Biviano+13,Guennou+14,MBM14,Balestra+16,Maughan+16,Biviano+17b,Sartoris+20} and for stacks of several clusters \citep[e.g.,][]{CYE97,vanderMarel+00,BG03,KBM04}. Results from these studies have been used to constrain the concentration--mass relation, in most cases by adopting the NFW model \citep{GGS16,Biviano+17a}. In general, these studies confirm that the NFW model provides an acceptable fit to the cluster $M(r)$, albeit with considerable variance from cluster to cluster, and there is a reasonable agreement between the observed concentration--mass relation and that predicted by $\Lambda$CDM cosmological numerical simulations.
However, all these studies involve clusters at $z<1$. There are only a few determinations of the $M(r)$ of clusters of galaxies at $z \sim 1$ or above, the one by \citet[][hereafter B+16]{Biviano+16} based on the phase-space distribution of cluster galaxies in the GCLASS sample \citep{Muzzin+12}, the ones by \citet{BDPV14} and \citet{AECS16} based on X-ray observations, and the one by \citet{SGEM15} based on gravitational lensing.

Previous observational determinations of $Q(r)$ and $\qr(r)$ for clusters of galaxies are those of \citet{MBM14} at $z=0.09$, \citet{Biviano+13} at $z=0.44$, and \citetalias{Biviano+16} at $z \sim 1$. All these studies found results that agree with the theoretical predictions, namely that $Q(r)$ and $\qr(r)$ follow power laws with the predicted slopes.

Observational determinations of $\beta(r)$ for clusters of galaxies are also in general agreement with predictions from numerical simulations \citep[e.g.,][]{Diaferio99,Munari+13,Lotz+19}, with $\beta(r) \simeq 0$ near the cluster center and increasing outside. This means cluster galaxies are on isotropic orbits near the center and increasingly radial orbits outside \citep[e.g.,][]{NK96,Mahdavi+99,BK04,BP09,Lemze+09,Biviano+13,MBM14,Annunziatella+16,Capasso+19,Mamon+19,SMH19}. There is, however, considerable variance in the shape of $\beta(r)$ from cluster to cluster \citep{HL08,Benatov+06,AADDV17}, as also found in cluster-sized halos from cosmological simulations \citep[][hereafter MBB]{MBB13}. At $z \sim 0$, determinations of cluster $\beta(r)$ indicate a dichotomy in the orbits of early-type, red, and quiescent galaxies on one side, and late-type, blue, star-forming galaxies, on the other. Early-type, red, quiescent galaxies move on isotropic orbits also at large distances from the cluster center, while late-type, blue, star-forming galaxies display more radially elongated orbits \citep[e.g.,][and references therein]{Mamon+19}. At higher-$z$, the orbits of red, quiescent galaxies are more similar to those of blue, star-forming galaxies, that is, more radial outside the center of the cluster \citepalias[e.g.,][and references therein]{Biviano+16}. 

In this paper we use the GOGREEN spectroscopic data set \citep{Balogh+17,Balogh+21} complemented with the GCLASS spectroscopic data set \citep{Muzzin+12} to probe in detail the internal dynamics of clusters of galaxies at an unprecedented high $z$. In particular, we determine the $M(r)$, $\beta(r)$, $Q(r)$,  and $\qr(r)$ of a stack of 14 clusters at redshift $0.87 \leq z \leq 1.37$ by using the MAMPOSSt method \citepalias{MBB13} and the Jeans equation inversion technique \citep{BM82,SSS90}. Our analysis is similar to that of \citetalias{Biviano+16}, which was limited to the GCLASS data set. As in \citetalias{Biviano+16} we consider the subsamples of star-forming and quiescent galaxies, and in addition we  consider subsamples of clusters in two bins of cluster $z$ and $\mvir$, and in two bins of galaxy stellar mass $M_{\star}$. With respect to our previous analysis, here we increase the number of clusters (from 10 to 14) and the total number of cluster members (from 418 to 581). 

The structure of this paper is the following. In Sect.~\ref{s:data} we describe our data set, how we identify cluster members (Sect.~\ref{ss:members}), the construction of the {\em ensemble} cluster from stacking (Sect.~\ref{ss:stack}), and the completeness of the spectroscopic sample (Sect.~\ref{ss:comp}). In Sect.~\ref{s:mass} we describe the methodology  we used to determine $M(r)$ (Sect.~\ref{ss:MAMmeth}), and provide the results of our analysis  for the $M(r)$ of the {\em ensemble} cluster and its subsamples. We compare the results with theoretical predictions and previous observational results (Sect.~\ref{ss:MAMres}). We also compare $\rho(r)$ with the number density and $M_{\star}$ density profiles of cluster members (Sect.~\ref{ss:MAMres}). In Sect.~\ref{s:beta} we describe the methodology by which we determine $\beta(r)$ (Sect.~\ref{ss:jeansinvmeth}) and provide the results of our analysis for the $\beta(r)$ of the {\em ensemble} cluster and its subsamples (Sect.~\ref{ss:jeansinvres}). In Sect.~\ref{s:PPSD} we determine the $Q(r)$ and $\qr(r)$ of the {\em ensemble} cluster and compare them to the theoretical predictions. In Sect.~\ref{s:disc} we discuss our results, separately for $M(r)$ (Sect.~\ref{ss:discmr}), $\beta(r)$ (Sect.~\ref{ss:discbr}), and $Q(r), \qr(r)$ (Sect.~\ref{ss:discqr}). In Sect.~\ref{s:conc} we provide our conclusions.

Throughout this paper we adopt the following cosmological parameter values: a Hubble constant $H_0=70\,{\rm km\,s}^{-1} {\rm Mpc}^{-1}$, a present-day matter density $\Omega_{\mathrm{m}}=0.3$, and a curvature parameter value $\Omega_{k}=0$.

\section{The data set}\label{s:data}
The cluster sample studied in this work is drawn from the GOGREEN \citep{Balogh+17,Balogh+21}
and GCLASS \citep{Muzzin+12} surveys. The GOGREEN survey targeted 27 clusters and groups at $1.0 < z < 1.5$. Three of these clusters were discovered by the  South Pole Telescope (SPT) survey \citep{Brodwin+10,Foley+11,Stalder+13}, and another nine were taken from the Spitzer Adaptation of
the Red-sequence Cluster Survey \citep[SpARCS,][]{Muzzin+09,Wilson+09,Demarco+10}. Nine groups in the COSMOS and Subaru-XMM Deep Survey \citep{Finoguenov+07,Finoguenov+10,George+11} complete the GOGREEN survey. The GCLASS survey targeted ten clusters at $0.8<z<1.3$, all taken from the SpARCS; five of them were also targeted in the GOGREEN survey.

The core observations of both GCLASS and GOGREEN were extensive multi-object spectroscopy of cluster galaxies with the GMOS spectrograph at the Gemini telescopes \citep{Hook+04}, with a spectroscopic resolving power $R=440$ and with a typical individual $cz$ uncertainty of 278 $\ks$, corresponding to a cluster rest-frame uncertainty $\delta_z <154 \, \ks$. This uncertainty is small compared to the $\slos$ of the GCLASS and GOGREEN clusters, and thus fully adequate for a dynamical analysis, given that $\delta_z$ contributes to the observed $\slos$ in quadrature \citep{HN79,DDZdT80}. The combined GCLASS and GOGREEN sample contains 2257 unique objects with good quality $z$ measurements \citep[as defined in][]{Balogh+21}: 1529 from GOGREEN and 728 from GCLASS. To this data set we add 112 unique objects with good $z$ measurements from the literature (references are given in Table~\ref{t:data}).

In this paper we only consider the 14 clusters with  $20$ spectroscopic members or more (membership determination is described in Sect.~\ref{ss:members}). Cluster velocity distributions based on fewer than $20$ members can be biased low by the effect of dynamical friction \citep{Old+13,Saro+13}, and cluster velocity dispersion estimates based on fewer than $20$ members have been shown to be statistically unreliable \citep{Girardi+93}. The list of the 14 clusters and their properties are given in  Table~\ref{t:data}.

\begin{table*}
\centering
\caption{List of cluster properties}
\label{t:data}
\resizebox{\textwidth}{!}{
  \begin{tabular}{ccrrrrrrrc}
  \hline
  Name & sample & $N_{\rm GG,GC}$ & ~~~~$N_{\rm lit}$ & $N_{\rm M}$ ($N_{\rm m}$) & $\overline{z}$~~~~~~~~~~~~~ & $\slos$~~~ & $\rvir$~~~~~~~~~~ & $\mvir$~~~ & Ref.s \\
    &   &   &  & & & [$\ks$] & [Mpc]~~~~~~~~        &   [$10^{14} \, {\rm M}_{\odot}$] &       \\
                \hline
  SpARCS0034 & GCLASS   &  67  &  -- & 31.5 (32) & $0.8673 \pm 0.0005$ & $ 405 \pm 51$  & $0.58, 0.58, 0.61$ & $ 0.6 \pm 0.2$ &   \\
  SpARCS0035 & GGGC     & 129  &  -- & 28.0 (29) & $1.3357 \pm 0.0013$ & $ 840 \pm 111$ & $0.93, 0.90, 1.01$ & $ 3.8 \pm 1.5$ &   \\
  SpARCS0036 & GCLASS   &  70  &  -- & 43.0 (48) & $0.8697 \pm 0.0008$ & $ 799 \pm 82$  & $1.12, 1.06, 1.21$ & $ 3.6 \pm 1.1$ &   \\
  SpARCS0215 & GCLASS   &  61  &  -- & 43.0 (44) & $1.0035 \pm 0.0007$ & $ 656 \pm 70$  & $0.85, 0.88, 0.89$ & $ 2.4 \pm 0.8$ &   \\
  SpARCS0335 & GOGREEN  &  66  &  67 & 23.0 (27) & $1.3690 \pm 0.0010$ & $ 542 \pm  75$ & $0.67, 0.69, 0.76$ & $ 1.8 \pm 0.7$ & a \\
  SpARCS1047 & GCLASS   &  68  &  -- & 29.0 (29) & $0.9566 \pm 0.0008$ & $ 668 \pm 89$  & $0.91, 0.91, 0.99$ & $ 2.5 \pm 1.0$ &   \\
  SpARCS1051 & GGGC     & 185  &  -- & 42.0 (42) & $1.0344 \pm 0.0007$ & $ 689 \pm  75$ & $0.88, 0.84, 0.95$ & $ 2.2 \pm 0.7$ &   \\
  SpARCS1613 & GCLASS   &  96  &  -- & 68.5 (86) & $0.8701 \pm 0.0009$ & $1185 \pm 90$  & $1.59, 1.54, 1.72$ & $11.1 \pm 2.5$ &   \\
  SpARCS1616 & GGGC     & 214  &  -- & 59.5 (60) & $1.1562 \pm 0.0007$ & $ 782 \pm 71$  & $0.92, 0.92, 1.00$ & $ 3.3 \pm 0.9$ &   \\
  SpARCS1634 & GGGC     & 190  &  -- & 63.0 (70) & $1.1780 \pm 0.0007$ & $ 715 \pm  60$ & $0.85, 0.85, 0.88$ & $ 2.7 \pm 0.7$ &   \\
  SpARCS1638 & GGGC     & 174  &  -- & 56.0 (56) & $1.1938 \pm 0.0006$ & $ 564 \pm  53$ & $0.71, 0.73, 0.74$ & $ 1.7 \pm 0.5$ &   \\
     SPT0205 & GOGREEN  &  65  &   5 & 28.0 (28) & $1.3227 \pm 0.0010$ & $ 678 \pm  91$ & $0.76, 0.85, 0.80$ & $ 3.1 \pm 1.2$ & b \\
     SPT0546 & GOGREEN  &  63  &  40 & 62.0 (67) & $1.0669 \pm 0.0009$ & $ 977 \pm  84$ & $1.17, 1.15, 1.22$ & $ 5.8 \pm 1.5$ & c \\
     SPT2106 & GOGREEN  &  67  &  14 & 43.0 (50) & $1.1307 \pm 0.0012$ & $1055 \pm 106$ & $1.23, 1.21, 1.24$ & $ 7.3 \pm 2.2$ & d \\
     \hline
  \end{tabular}}
\tablefoot{GOGREEN, GGGC, and GCLASS identify the data base: GOGREEN-only data, GOGREEN and GCLASS data, and GCLASS-only data, respectively. $N_{\rm GG,GC}$ is the number of objects with good-quality redshifts from either GOGREEN or GCLASS, and $N_{\rm lit}$ the number of objects with good-quality redshifts available in the literature for galaxies that do not have redshifts from either GOGREEN or GCLASS. The total number of galaxies with redshifts $N_{\rm tot}$ is obtained by the sum $N_{\rm GG,GC}+N_{\rm lit}$.
The references to the literature data are given in the last column: a: \citet{Nantais+16}, b: \citet{Stalder+13}, c: \citet{Sifon+16}, d: \citet{Foley+11}. $N_{\rm M}$ is the weighted number of members among the $N_{\rm tot}$ objects (see text).  In parentheses we list $N_{\rm m}$, the number of galaxies that would be considered as members by at least one of the two cluster membership algorithms used in this analysis (see text for more details). $\overline{z}$ and $\slos$ are the weighted estimates of the cluster mean redshift and velocity dispersion, obtained by using the galaxy membership weights (1$\sigma$ errors are listed). The $\rvir$ column lists three estimates for each cluster, obtained respectively by the Clean procedure ($r_{200,C}$), and by MAMPOSSt ($r_{200,Mt}$ and $r_{200,Mc}$) using two different mass and anisotropy profiles (see text for more details). Uncertainties on $\rvir$ are not listed as they are proportional to the fractional uncertainties on $\slos$. The $\mvir$ column lists the values obtained from $r_{200,Mt}$, with uncertainties estimated as 3 times the fractional error on $\slos$.}
\end{table*}

\subsection{Cluster membership}
\label{ss:members}
To identify cluster members we proceeded in three steps. First we identified the main cluster peak in $z$ space \citep[following][]{Beers+91,Girardi+93} by selecting those galaxies in the cluster field with ${c \mid z-z_c \mid \leq 6000 \, \ks}$, where $c$ is the speed of light and $z_c$ is the initial cluster redshift estimate, taken from \citet{Balogh+17} for the GOGREEN clusters and \citetalias{Biviano+16} for the GCLASS clusters.

We then applied the Kernel Mixture Model (KMM) algorithm \citep{McLB88,ABZ94} to the distribution of redshifts located in the main peak. KMM estimates the probability that the $z$ distribution is better represented by $k$ Gaussians rather than a single one. Given the limited number of galaxies with redshifts available for each cluster, we only considered the case $k=2$.  This part of the procedure is meant to identify cases of merging subclusters close to the line of sight, and to separate these subcluster components from the main cluster. Secondary peaks are identified only in two clusters, SpARCS1051 and SpARCS1616. These secondary peaks correspond to two small groups of four galaxies each, and they are removed from the samples of cluster members. Overall, the KMM analysis suggests that the GOGREEN clusters have no strong contamination from fore- or background groups nor are they undergoing major mergers along the line of  sight.

The membership selection of the galaxies that are left in the sample
after the main-peak and KMM selections was then refined by using
two methods:
Clean \citepalias{MBB13} and CLUMPS, a new algorithm that we describe in detail in Appendix~\ref{s:CLUMPS}.
Both Clean and CLUMPS identify cluster members based on  location in projected phase-space $R, \vrf$, where $R$ is the projected radial distance from the cluster center and ${\vrf \equiv c \, (z-\overline{z})/(1+\overline{z})}$ is the rest-frame velocity, with the mean cluster redshift $\overline{z}$ estimated on the identified cluster members (see below). For the cluster center we use the position of the brightest cluster galaxy \citep[BCG;][]{vanderBurg+20}.

These two methods for cluster membership are conceptually very different. Clean is theoretically motivated, its parameters fixed by our understanding of the properties of cluster-sized halos extracted from cosmological numerical simulations. CLUMPS is instead less model-dependent, and is based only on the fact that a cluster of galaxies is a concentration in projected phase-space. Clean ensures stability thanks to the use of theoretically motivated models, but its results depend on how close  the chosen models are to reality. CLUMPS, on the other hand, depends on the choice of parameters that are not linked to a specific model,  thus  chosen with considerable freedom, and this can cause instability in the results. The combined use of Clean and CLUMPS helps in reducing possible systematics in the process of membership selection \citep{Wojtak+07} as we expect the two methods to compensate each other for their relative weaknesses.

More specifically, the Clean method uses an estimate of the cluster line-of-sight velocity dispersion, $\slos$, to provide a first estimate of the cluster mass from a scaling relation. It then adopts the NFW profile, a theoretical concentration--mass relation \citep{MDvdB08}, and a velocity anisotropy profile model \citep{MBM10}, to predict $\slos(R)$ and to iteratively reject galaxies with ${\mid \vrf \mid > 2.7 \, \slos}$ at any radius \citepalias[for more details, see][]{MBB13}. The adopted profiles are known to be a good description of real cluster profiles on average. Adopting different models would not make a large difference in the membership selection since they would predict similar velocity dispersion profiles
(for an example, see Fig.~\ref{f:vdpmamp}), but the adopted profiles might not be a good description of the individual cluster profile. For this reason we supplement Clean with CLUMPS. CLUMPS evaluates the density of galaxies in projected phase-space, by counting galaxies in bins of $R$ and $\vrf$, and applies a convolution of this density distribution with a Gaussian filter in Fourier space. Then for each bin in the radial direction CLUMPS identifies the main peak in velocity space. The minima around this peak define the velocity boundaries outside which interlopers are removed in any given bin (see Appendix~\ref{s:CLUMPS} for more details).

We ran each of these two methods twice. On the first run
we defined $\overline{z}$ as the average redshift of the galaxies that
passed the main-peak and KMM selection. On the second run we re-defined
$\overline{z}$ as the average redshift of the galaxies selected as
members in the first run.

In Col.~5 of Table~\ref{t:data} we list $N_{\rm{M}}$, the weighted number of
cluster members, assigning weight=1.0 to objects selected as members by both membership algorithms (described in the text), weight=0.5 to objects selected as members by only one of the two membership algorithms, and weight=0.0 to objects that are not selected as members by either of the two algorithms. In the same Col.~5 we also list in parentheses $N_{\rm{m}}$, the number of galaxies identified as members by either of the two membership algorithms.
In Cols.~6 and 7  we provide the cluster mean
redshift $\overline{z}$ and, respectively, velocity dispersion $\slos$,
obtained on the sample of $N_{\rm{M}}$ members with the weighted mean
and weighted dispersion estimators, using the galaxy membership weights. 

\subsection{The ensemble cluster}
\label{ss:stack}
We did not have a sufficient number of spectroscopically identified member galaxies per cluster (typically $\gtrsim 200$) to allow the determination of the mass and anisotropy profile of each cluster independently. To improve the statistics we built an {\em ensemble} cluster by stacking the data of our 14 clusters, a procedure widely used in the literature \citep[e.g.,][]{Carlberg+97-equil,BG03,MG04,KBM04,Rines+13,Cava+17}. We based our stacking procedure on the theoretical expectation that the mass profile of massive halos are almost homologous. Both theory and observations indicate that, on average, the $\mr$ of clusters have very similar concentrations at different redshifts and masses \citep[e.g.,][]{DeBoni+13,Biviano+16}, and therefore depend on a single parameter, $\rvir$. We could then build the {\em ensemble} cluster by stacking together all the members of our 14 clusters in projected phase-space, by normalizing the clustercentric distances $R$ by $\rvir$ and the rest-frame velocities $\vrf$ by $\vvir$. We therefore needed to estimate $\rvir$ for each of our 14 clusters. To check for possible systematic errors, we 
provided three estimates of $\rvir$ of each individual cluster. These three estimates are listed in
Col.~8 of Table~\ref{t:data}.

The first estimate is a by-product of the Clean procedure. Clean does in fact use $\slos$ to predict the cluster $\rvir$ in an iterative way \citepalias[see Appendix B of][]{MBB13}. We call this $\rvir$ estimate $r_{200,C}$. The second $\rvir$ estimate, called $r_{200,Mt}$, comes from application of the MAMPOSSt method (described in Sect.~\ref{ss:MAMmeth}) to the sample of members defined by Clean and CLUMPS, using membership weights. Since the individual cluster data sets are not large enough to  constrain several parameters of the mass and anisotropy profiles, we only allowed the value of $\rvir$ as a free parameter in the MAMPOSSt procedure. We assumed the light-traces-mass hypothesis, by forcing the galaxy number density profile to have the same shape as the mass density profile, and fixed the latter to the NFW profile with a concentration $\cvir=3.5$, a typical value for clusters at $z \sim 1$ according to cosmological numerical simulations \citep[e.g.,][]{Duffy+08}. We also assumed the $\br$ model of \citet{Tiret+07}, $\br=\beta_{\infty} \, r/(r+\rtwo)$, where  $\rtwo$ is the radius where the logarithmic derivative of the mass density profile equals $-2$, and we took $\beta_{\infty}=0.5$. This $\br$ model was found to provide a good fit to the $\br$ of simulated clusters \citep[e.g.,][]{MBM10}. Like the previous one,  our third $\rvir$ estimate also comes from an application of the MAMPOSSt method. Here we do not use theoretical estimates of $\mr$ and $\br$; instead, we adopt the best-fit profiles obtained by \citetalias{Biviano+16} on the GCLASS sample of clusters, namely a Burkert profile \citep{Burkert95}  for $\mr$ and an Opposite profile \citep{Biviano+13}  for $\br$. We call this   $\rvir$ estimate $r_{200,Mp}$.

\begin{table*}
\centering
\caption{Properties of the {\em ensemble} cluster and its subsamples}
\label{t:ensemble}
\begin{tabular}{lrrrrrrr}
  \hline
  Sample & $N_c$~  & \multicolumn{2}{c}{$N_M (N_m)$} & $f_{\rm{SF}}$ & \multicolumn{1}{c}{$z$} & \multicolumn{1}{c}{$\rvir$} & \multicolumn{1}{c}{$\mvir$} \\
         &        & [All radii]~~~ & $[0.05 \leq R \leq \rvir]$ &    &        & \multicolumn{1}{c}{[Mpc]} & \multicolumn{1}{c}{[$10^{14} \, {\rm M}_{\odot}$]} \\
\hline      
       {\em ensemble} & 14 &  527.5 ( 581) &  420.5 ( 467) & 0.25 & $ 1.07 \pm 0.05$ & $0.98 \pm 0.07$ & $3.6 \pm 0.8$ \\ 
 {\em low-}$\mvir$ &  7 &  231.5 ( 256) &  162.0 ( 181) & 0.33 & $ 1.12 \pm 0.07$ & $0.78 \pm 0.04$ & $2.0 \pm 0.3$ \\ 
{\em high-}$\mvir$ &  7 &  296.0 ( 325) &  258.5 ( 286) & 0.21 & $ 1.04 \pm 0.07$ & $1.14 \pm 0.09$ & $5.5 \pm 1.3$ \\ 
     {\em low-}$z$ &  7 &  281.5 ( 306) &  242.5 ( 265) & 0.23 & $ 0.96 \pm 0.03$ & $1.07 \pm 0.12$ & $4.1 \pm 1.4$ \\ 
    {\em high-}$z$ &  7 &  246.0 ( 275) &  178.0 ( 202) & 0.29 & $ 1.21 \pm 0.04$ & $0.89 \pm 0.07$ & $3.1 \pm 0.7$ \\ 
\hline
\end{tabular}
\tablefoot{$N_c$ is the number of clusters in the sample. $N_M$ and $N_m$ are respectively the sum of the membership weights and the total number of members with $M_{\star} \geq 10^{9.5} \, {\rm M}_{\odot}$ in the clusters that are part of the sample, at all radii (Col.~3) and in the radial range used for the dynamical analysis (Col.~5). $f_{\rm SF}$ is the fraction of star-forming galaxies among the members in the $0.05 \leq R \leq \rvir$ radial range, identified based on their $U-V$ and $V-J$ rest-frame colors \citep[for more details see text and][]{vanderBurg+20}.
$z$ and  $\rvir$ are evaluated using $N_M$-weighted averages of the corresponding quantities for the clusters that are part of the sample. $\mvir$ is derived from $\rvir$, given $z$ (see footnote \ref{foot:mvir}).} 
\end{table*}

\begin{figure}
\centering
\includegraphics[width=.5\textwidth]{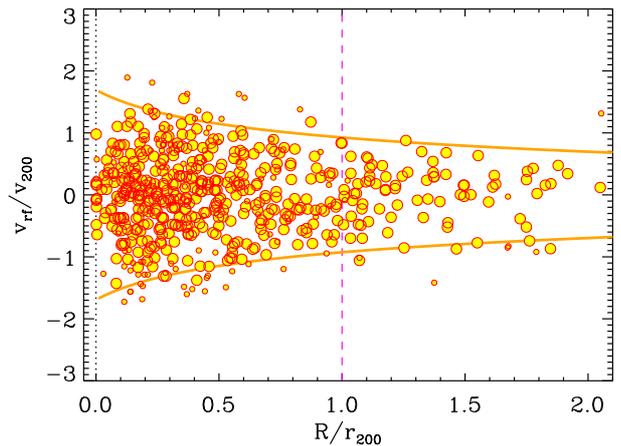}
\caption{Normalized projected phase-space diagram of the {\em ensemble} cluster obtained using 
$\rvir=r_{200,Mt}$. Large dots have membership weights of 1.0, small dots membership weights of 0.5. Escape-velocity curves are shown based on a NFW $\mr$ with $\cvir=3.5$ and with a Tiret $\br$ with $\beta_{\infty}=0.5$.} 
\label{f:npps}
\end{figure}

The  three $\rvir$ estimates are rather similar; on average, $r_{200,C}$ and $r_{200,Mt}$ are equal, $r_{200,C}$ and $r_{200,Mp}$ differ by 6\%, and $r_{200,Mt}$ and $r_{200,Mp}$ differ by 6\%. The maximum difference found among the three $\rvir$ estimates for any of our 14 clusters is 13\%.
We adopted $\rvir=r_{200,Mt}$ to build our {\em ensemble} cluster in normalized projected phase-space, $R/\rvir$ and $\vrf/\vvir$. We  verified that our conclusions are not sensitive to this choice. The normalized projected phase-space distribution obtained using $\rvir=r_{200,Mt}$ is shown in Fig.~\ref{f:npps}. 

To increase the spectroscopic completeness of the sample (important for the dynamical analysis; see Sect.~\ref{ss:comp}) we restrict the dynamical analysis to galaxies in the {\em ensemble} clusters with stellar mass $M_{*} \geq 10^{9.5} {\rm M}_{\odot}$. This is the $M_{*}$ limit to which the exposure times, and the associated depths, of the GOGREEN photometry were tailored \citep{vanderBurg+20}. This $M_{*}$ selection only removes 7\% of the cluster members.

Properties of this {\em ensemble} cluster (and subsamples extracted from it) are listed in Table~\ref{t:ensemble}.
They are evaluated as weighted averages of the corresponding properties of the 14 clusters that make the {\em ensemble} cluster, using the number of members of each cluster as a weight.
We considered the following subsamples of the {\em ensemble} sample  in the current analysis:
\begin{itemize}
    \item {\em low-}$\mvir$, 7 clusters with $\mvir$ below the median $\mvir$ of the {\em ensemble} sample, $\mvir=2.8 \times 10^{14} \, {\rm M}_{\odot}$;
    \item {\em high-}$\mvir$, 7 clusters with $\mvir$ above the median $\mvir$ of the {\em ensemble} sample;
    \item {\em low-}$z$, 7 clusters with $z$ below the median $z$ of the {\em ensemble} sample, $z=1.10$;
    \item {\em high-}$z$, 7 clusters with $z$ above the median $z$ of the {\em ensemble} sample.
\end{itemize}

\subsection{Spectroscopic completeness}
\label{ss:comp}
Our analysis is impacted by the different survey strategies, and particularly by which galaxies were targeted spectroscopically. Spectroscopic incompleteness is not likely to affect the observed distribution of cluster galaxy velocities since velocity segregation with galaxy mass has a very small effect in clusters \citep{BGMM92,ABM98,Biviano+02,Goto05}. However, since the number  density profiles are affected by spectroscopic incompleteness, it is essential to correct for it in the Jeans dynamical analysis. More precisely, the global value of spectroscopic incompleteness does not affect the dynamical analysis because the number density profile only enters the Jeans equation through its logarithmic derivative. What matters is the radial dependence of the spectroscopic incompleteness. If a sample has different levels of completeness at different distances from the cluster center, the inferred number density profile will differ from the intrinsic one. 

It is not feasible to evaluate the spectroscopic correction as a function of $R/\rvir$ 
for each individual cluster in a robust way because of poor statistics. 
However, the spectroscopic completeness (including the dependence on stellar mass and radial distance) is very different for the GOGREEN and GCLASS surveys.  Therefore, to evaluate the spectroscopic completeness of our {\em ensemble} cluster, we 
divided our {\em ensemble} cluster into three subsamples: one with only GOGREEN spectroscopic data, another with both GOGREEN and GCLASS spectroscopic data, and a third  with only GCLASS spectroscopic data (clusters belonging respectively to the GOGREEN, GGGC, and GCLASS samples in Table~\ref{t:data}). We determined a spectroscopic completeness correction separately for the three subsamples of the {\em ensemble} cluster. 

We evaluated the spectroscopic completeness of each of the three subsamples following \citet{vanderBurg+20}. Specifically, in each of several radial bins, we counted the number of galaxies, $N_{\rm{p}}$, with photometric redshifts within a given range of the mean cluster redshift\footnote{Changing the $\Delta z_{\rm{p}}$ range from 0.08, the value used in \citet{vanderBurg+20}, to 0.12 does not modify the completeness curves in a significant way.}, $\mid \Delta z_{\rm{p}} \mid \leq 0.08$, and the number of these galaxies with good quality redshift measurements,
$N_{\rm{s}}$. Photometric redshifts were evaluated with the code \texttt{EAZY} of \citet{BvDC08}, as described in \citet{vanderBurg+20}.
We considered radial bins in units of $\rvir$, and only galaxies with stellar mass $M_{*} \geq 10^{9.5} {\rm M}_{\odot}$ (see Sect.~\ref{ss:stack}). In the same radial bins we estimated the fraction $f_M$ of the covered area that is masked because of the presence of bright stars, diffraction spikes, and reflective halos, or because of artifacts in any photometric band \citep[see][for details]{vanderBurg+20}. The spectroscopic completeness in each radial bin was then evaluated as ${C = (1-f_M) \, N_{\rm{s}}/N_{\rm{p}}}$.

\begin{figure}
\centering
\includegraphics[width=.5\textwidth]{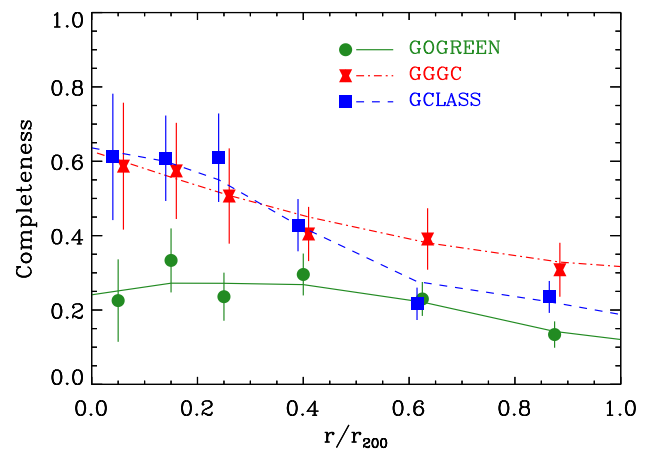} 
\caption{Points with 1$\sigma$ error bars: radial spectroscopic completeness profiles for galaxies with $M_{*} \geq 10^{9.5} {\rm M}_{\odot}$ in three subsamples of the {\em ensemble} cluster: GOGREEN (green dots), GCLASS (blue squares), and GGGC (red crosses). The symbols representing the three subsamples have been slightly displaced along the abscissa for display purposes. Solid, dashed, and dash-dotted curves represent LOWESS smoothing of the data represented by the dots, squares, and crosses, respectively.}
\label{f:comp}
\end{figure}

We plot in Fig.~\ref{f:comp} the spectroscopic completeness profiles for the three subsamples of the {\em ensemble} cluster, in the radial range that is adopted for the fit of an NFW model to the number density profile (see Sect.~\ref{ss:MAMmeth}). To avoid over-interpreting completeness oscillations due to Poisson uncertainties, we smoothed these profiles with the LOWESS technique \citep[e.g.,][]{Gebhardt+94} and adopted the smoothing curves to correct the observed number density profiles for incompleteness.

\section{The mass profile}\label{s:mass}
\subsection{Methodology: MAMPOSSt}\label{ss:MAMmeth}
The MAMPOSSt method \citepalias{MBB13} performs a maximum likelihood fit of the $\mr$ and $\br$ models to the projected phase-space distribution of cluster members. MAMPOSSt assumes a shape for the 3D velocity distribution (a Gaussian in the current implementation of the method). It has been successfully tested on simulated halos from cosmological simulations that included both dynamically relaxed and unrelaxed halos, and it has already been applied to several data sets \citep[e.g.,][]{Biviano+13,Guennou+14,MBM14,Verdugo+16,Biviano+17a,Pizzuti+17,Mamon+19,Sartoris+20}. 

The MAMPOSSt method is based on the Jeans equation of dynamical equilibrium \citep{BT87,Pratt+19}, and therefore it should be applied to clusters in dynamical equilibrium. We cannot exclude the possibility that some of the 14 clusters forming the {\em ensemble} cluster are not in dynamical equilibrium. However, there is no evidence for ongoing major mergers from the KMM analysis (see Sect.~\ref{ss:members}), and we limited our analysis to $R \leq \rvir$ to exclude unvirialized regions as completely as possible from our analysis\footnote{We checked that further restricting the radial range to $R \leq 0.8 \, \rvir$ (corresponding roughly to an overdensity $\Delta=300$) does not change our $M(r)$ in a significant way.}.
We also exclude the very inner region, $R \leq 0.05$ Mpc, where the gravitational potential of the BCG becomes dominant relative to the gravitational potential of the cluster as a whole \citep[see, e.g.,][]{BS06}\footnote{It is possible to extend MAMPOSSt by adding the BCG stellar kinematics as a constraint \citep{Sartoris+20}, but a precise determination of the velocity dispersion profile of the BCG is required, and it is not available for any of our 14 clusters.}. 

MAMPOSSt assumes spherical symmetry. While clusters are not spherically symmetric systems \citep[e.g.,][]{Chiu+18}, the {\em ensemble} cluster is spherically symmetric by construction unless the clusters that compose it have been selected with a preferential orientation along the line of sight \citep[see also Sect.~4 in][for a discussion of the spherical assumption in the case of an {\em ensemble} cluster]{vanderMarel+00}.

We chose to run MAMPOSSt in the Split mode \citepalias[see Sect. 3.4 in][]{MBB13}. We first determined the best fit to the number density profile, $\nu(r)$, of the {\em ensemble} cluster members, and then we determined the best fit to the distribution of cluster member velocities as a function of their radial positions. We used the projected NFW model \citep{Bartelmann96} to fit the projected number density profile, $N(R)$, with a maximum likelihood technique. The only free parameter is the scale radius $\rnu/\rvir$ where the logarithmic derivative of the number density profile equals $-2$, or, equivalently, $c_{\nu}=\rvir/\rnu$.
In the maximum likelihood fitting technique, the normalization of the fitting function is not a free parameter, as it is constrained by the requirement that the total (completeness corrected) number of observed galaxies is identical to the integral of the fitting function over the radial range of the fit \citep{Sarazin80}. In the fitting procedure we used the spectroscopic radial completeness curves shown in Fig.~\ref{f:comp} to correct for incompleteness by weighting each galaxy by $1/C_j(R_i),$ where $C_j(R_i)$ is the value of the completeness curve at the radial position of the galaxy, $R_i$, and $j$ identifies the subsample of clusters the galaxy belongs to, GOGREEN, GGGC, or GCLASS (see Sect.~\ref{ss:comp} and Fig.~\ref{f:comp}). In the fit we also weighted for membership, so that the total weight is $m_i/C_j(R_i)$, where $m_i=1.0$ when the galaxy is classified as a cluster member by both the Clean and the CLUMPS algorithms, and $m_i=0.5$ when it is classified as a cluster member by only one of the two algorithms.

While not needed for MAMPOSSt, we also determined the fit of an NFW model to the stellar mass density profile, by weighting each galaxy by $M_{\star,i} \, m_i/C_j(R_i)$. We did this to compare the stellar mass concentration to the total mass concentration (see Sect.~\ref{ss:MAMres}).

There are no limitations in the number of parameters that can be used to describe $\mr$ and $\br$ in  MAMPOSSt. Given the size of our data set we considered only three free parameters, two to describe $\mr$ and one to describe $\br$. For $\mr$ the free parameters we considered are $\rvir$ and $\rtwo$. The ratio $\cvir=\rvir/\rtwo$ defines the concentration, and therefore the shape of the mass profile, while $\rvir$ is related to $\mvir$ and this defines the $\mr$ normalization. Given that we were working on an {\em ensemble} cluster, where radii are in units of $\rvir$ and velocities in units of $\vvir$, rather than allowing $\rvir$ to be a free parameter, we also ran MAMPOSSt by fixing $\rvir$ to the weighted average value (listed in Table~\ref{t:ensemble})
of the 14 cluster $\rvir$ used to stack the clusters. 

We considered the following $\mr$ models (see Eqs. (4--6) in \citealt{Biviano+13} and Eq. (5) in \citealt{Sartoris+20} for the analytic expressions of these models): 
\begin{itemize}
\item the Burkert model \citep{Burkert95};
\item the Einasto model \citep{Einasto65,MBM10}, with $m=5$ \citep[as in][]{Biviano+13}; \item the Hernquist model \citep{Hernquist90}; 
\item gNFW, the generalized NFW model, which differs from the NFW model because the inner slope of the mass density profile, $\gamma$, is allowed to be different from 1. The characteristics of our data set do not allow us to constrain $\gamma$ as a free parameter. We chose instead to consider four different values, $\gamma=0.0, 0.5, 1.0, 1.5$, with $\gamma=1.0$ corresponding to the traditional NFW model. 
\end{itemize}

For $\br$, the only free parameter of MAMPOSSt depends on the chosen model. We considered the following $\br$ models:
\begin{itemize}
\item C: Constant anisotropy with radius, $\br=\beta_C$;
\item O: anisotropy of opposite sign at the center and at large radii \citep{Biviano+13},
$\br=\beta_{\infty} \, (r-\rtwo)/(r+\rtwo)$;
\item OM: from \citet{Osipkov79} and \citet{Merritt85-df}, rapidly increasing radial anisotropy with radius,
$\br= r^2/(r^2+r_{\beta}^2)$;
\item T: from \citet{Tiret+07}, increasing radial anisotropy with radius,
$\br=\beta_{\infty} \, r/(r+\rtwo)$;
\item Tl: see \citet{Tiret+07} and \citet{Mamon+19}, similar to the T model, but using the scale radius of the galaxy distribution rather than the scale radius of the mass distribution,
$\br=\beta_{\infty} \, r/(r+\rnu)$.
\end{itemize}
Model C   has been frequently used in previous studies because of its simplicity \citep[e.g.,][]{Merritt87,vanderMarel+00,LM03}, but models T and Tl  appear to better represent the $\br$ of cosmological halos \citep[e.g.,][]{MBM10,MBB13,Munari+13}. Model O (respectively OM)   provides the best fit to the $\br$ of GCLASS clusters \citepalias{Biviano+16} \citep[respectively EDisCS clusters;][]{BP09}. 

In total we have $7 \times 5=35$ MAMPOSSt runs for each combination of the 7 $\mr$ and 5 $\br$ models. We find the best-fit parameters using the NEWUOA software for unconstrained optimization \citep{Powell06}. Using the Bayes information criterion \citep[BIC,][]{Schwarz78} we find that the results obtained by fixing $\rvir$ to the weighted average of the 14 cluster $\rvir$ values are statistically favored compared to the results obtained by leaving $\rvir$ as a free parameter. In the following we therefore present only the results obtained with two free parameters, $\rtwo$ in units of $\rvir$ or, equivalently, $\cvir$, for $\mr$; and $\beta_C$, $r_{\beta}$, or $\beta_{\infty}$, depending on the chosen $\br$ model.

\subsection{Results}\label{ss:MAMres}
\begin{figure}
\centering
\includegraphics[width=.5\textwidth]{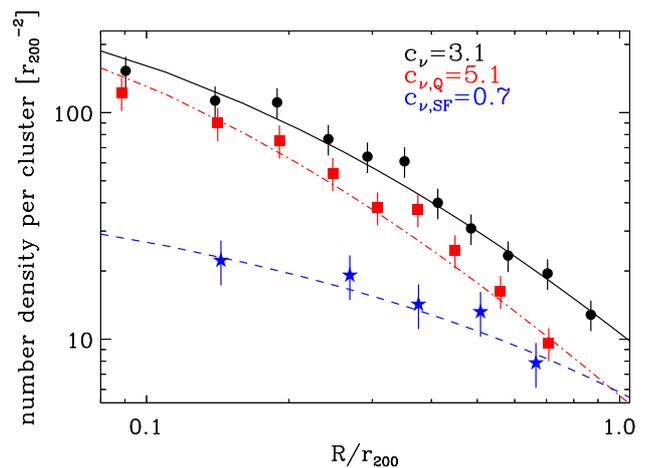} 
\caption{Projected galaxy number density profile of the {\em ensemble} cluster (dots with 1$\sigma$ Poissonian error bars), corrected for spectroscopic incompleteness, and best fit with the NFW model (curve). Black dots and solid curve: all member galaxies; red squares and dash-dotted curve: quiescent galaxies; blue stars and dashed curve: star-forming galaxies. Best-fit concentration values are listed.}
\label{f:densprof}
\end{figure} 

\begin{table*}
\centering
\caption{Concentrations of the {\em ensemble} cluster and its subsamples}
\label{t:concs}
\begin{tabular}{lrrrrr}
  \hline
  Sample & \multicolumn{1}{c}{$c_{\nu}$} & \multicolumn{1}{c}{$c_{\nu,\rm{Q}}$} & \multicolumn{1}{c}{$c_{\nu,\rm{SF}}$} &\multicolumn{1}{c}{$c_{\star}$} & \multicolumn{1}{c}{$\cvir$} \\
\hline
      {\em ensemble} & $ 3.1 \pm 0.5$ & $ 5.1 \pm 0.9$ & $ 0.7 \pm 0.4$ & $ 4.1 \pm 0.7$ & $ 3.2 \pm (0.23,0.12,0.05)$ \\ 
   {\em low-}$\mvir$ & $ 2.0 \pm 0.6$ & $ 4.7 \pm 1.7$ & $ 0.3 \pm 0.4$ & $ 2.3 \pm 0.7$ & $ 3.5 \pm (0.20,0.18,0.24)$ \\ 
  {\em high-}$\mvir$ & $ 4.2 \pm 0.8$ & $ 6.0 \pm 1.4$ & $ 1.2 \pm 0.7$ & $ 6.1 \pm 1.3$ & $ 3.1 \pm (0.27,0.15,0.09)$ \\ 
       {\em low-}$z$ & $ 3.2 \pm 0.7$ & $ 5.8 \pm 1.5$ & $ 0.3 \pm 0.4$ & $ 2.0 \pm 0.5$ & $ 3.7 \pm (0.43,0.22,0.21)$ \\ 
      {\em high-}$z$ & $ 3.2 \pm 0.8$ & $ 5.4 \pm 1.7$ & $ 1.5 \pm 1.0$ & $ 4.6 \pm 1.2$ & $ 2.7 \pm (0.22,0.17,0.10)$ \\ 
      \hline
\end{tabular}
\tablefoot{$c_{\nu}, c_{\nu,\rm{Q}}$, and $c_{\nu,\rm{SF}}$ are the concentrations of the best-fit NFW models to the number density profiles of all, quiescent, and star-forming galaxies, respectively. $c_{\star}$ is the concentration of the best-fit NFW model to the $M_{\star}$ density profile. $\cvir$ is the ${\cal L}$-weighted average of the concentrations of the 35 best-fit model $M(r)$ obtained via the MAMPOSSt analysis. The three components to the total uncertainty on $\cvir$ are given in parentheses:  $\delta_{c,s}, \delta_{c,\nu},$ and $\delta_{c,j}$, respectively (see Sect.~\ref{ss:MAMres} for their definitions).}
\end{table*}

In Fig.~\ref{f:densprof} we show $N(R)$ of
the {\em ensemble} sample corrected for spectroscopic incompleteness,
and the best-fit (projected) NFW model, for all member galaxies, and  for the subsamples of quiescent and star-forming galaxies. Following \citet{vanderBurg+20}, we identify quiescent and star-forming galaxies based on their $U-V$ and $V-J$ rest-frame colors, with quiescent galaxies satisfying the following criteria:
\begin{equation}
 U-V > 1.3 \,\,\,\cap\,\,\, V-J < 1.6 \,\,\,\cap\,\,\, U-V > 0.60+(V-J). 
\end{equation}  
We list in Table~\ref{t:concs} the best-fit values of $c_{\nu}$ and their uncertainties for the 
{\em ensemble} cluster
and all the subsamples extracted from it.

\begin{figure}
\centering
\includegraphics[width=.5\textwidth]{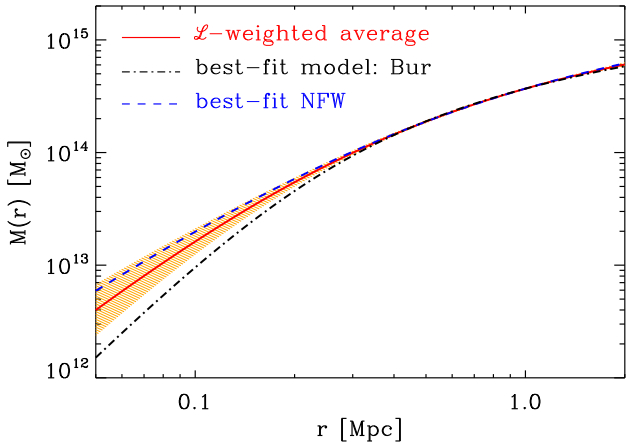} 
\caption{MAMPOSSt results for the $\mr$ of the {\em ensemble} cluster. Red curve and orange shading: $\cal{L}$-weighted average of all 35 best-fit model $\mr$ from the MAMPOSSt analysis, and 1$\sigma$ error. Blue dashed curve: highest-$\cal{L}$ among NFW models. Black dot-dashed curve: highest-$\cal{L}$ model (Burkert).}
\label{f:mr}
\end{figure}

In Fig.~\ref{f:mr} we show the results of the MAMPOSSt analysis for $\mr$. In particular, we display the weighted average of all 35 best-fit model $\mr$, using the MAMPOSSt likelihoods $\cal{L}$ as weights; the best-fit NFW model; and the model with the highest likelihood, a Burkert $\mr$. Taken at face value, these results suggest that the preferred solutions have $\rho(r)$ with inner slopes that are less steep than in the NFW model. A similar result was suggested by \citetalias{Biviano+16} based on the GCLASS data.
However, the evidence is not statistically significant since all models provide
statistically acceptable fits to the data, according to
the likelihood-ratio test \citep{Meyer+75}. 

\begin{figure}
\centering
\includegraphics[width=.5\textwidth]{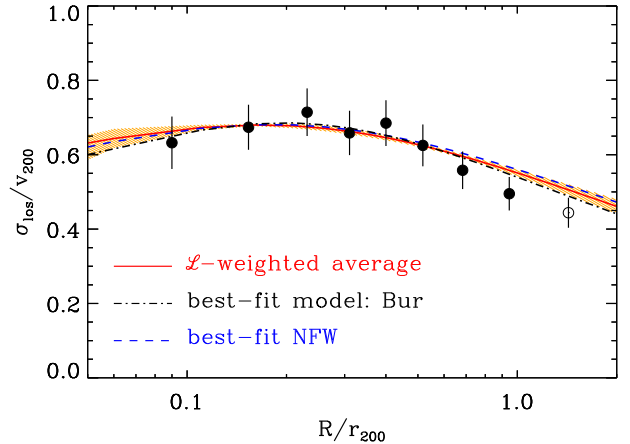} 
\caption{Line-of-sight velocity dispersion profile of the {\em ensemble} cluster (dots and 1$\sigma$ error bars, estimated using eq.~(16) in \citealt{BFG90}). Also shown are the MAMPOSSt results for $\mr$ and $\br$ projected onto the velocity dispersion profile. Red curve and orange shading: $\cal{L}$-weighted average of all 35 best-fit model $\mr$ from the MAMPOSSt analysis, and 1$\sigma$ error (reduced $\chi^2=0.6$). Blue dashed curve: highest-$\cal{L}$ among NFW models (reduced $\chi^2=0.7$). Black dot-dashed curve: highest-$\cal{L}$ model (Burkert, reduced $\chi^2=0.4$). The last point (empty dot) represents the velocity dispersion of galaxies that were not used in the MAMPOSSt analysis because they are at $R \geq \rvir$.}
\label{f:vdpmamp}
\end{figure} 

In general, there is no guarantee that a maximum likelihood solution also provides a good fit to the data if the chosen models are a poor choice. To allow a direct comparison with the data we project the $\mr+\br$ solutions of MAMPOSSt on the velocity dispersion profile and find very good agreement (see Fig.~\ref{f:vdpmamp}). Different models fit almost equally well to the observed velocity dispersion profile, a confirmation that we are unable to distinguish different mass profile models with the current data set.

We evaluate a $\cal{L}$-weighted average $\cvir \equiv \rvir/\rtwo$ from the 35 MAMPOSSt runs for each (sub)sample (see Table~\ref{t:concs}). The 1$\sigma$ errors on $\cvir$ come from the addition of three sources of errors, $\delta_c=\delta_{c,s}+\delta_{c,\nu}+\delta_{c,j}$, listed separately in Table~\ref{t:concs}. The first term, $\delta_{c,s}$, is the statistical error on the weighted average. The second term, $\delta_{c,\nu}$, accounts for our incomplete knowledge of $\nu(r)$. It is half the difference between the two values of $\cvir$ obtained by running MAMPOSSt with either the upper or the lower limit on $c_{\nu}$ as input parameter; these limits are obtained in the NFW maximum likelihood fit to $N(R)$. Finally, the third term, $\delta_{c,j}$, accounts for sample variance. It is estimated by a jackknife technique \citep{Efron82}, by taking the rms of the $\cvir$ obtained by running MAMPOSSt $N_c$ times on $N_c-1$ clusters, eliminating one different cluster at each MAMPOSSt run. Sample variance is not a dominant source of error in our analysis since the amplitude of $\delta_{c,j}$ is similar to the amplitudes of the other two error terms.

\begin{figure}
\centering
\includegraphics[width=.5\textwidth]{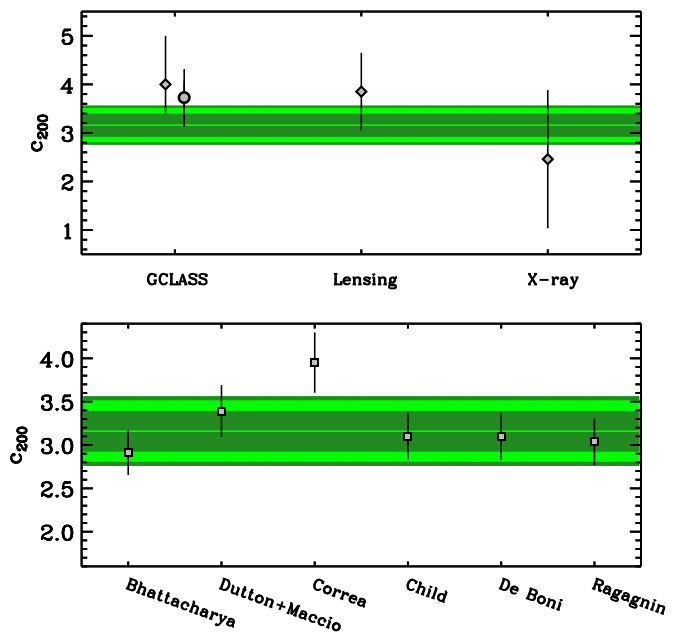} 
\caption{The concentration of the {\em ensemble} cluster compared to previous observational results and theoretical predictions. {\em Top panel:} Green horizontal line: $\cal{L}$-weighted average $\cvir$ for the {\em ensemble} cluster. Shadings represent the 1$\sigma$ error bar on the $\cvir$, with contributions from the three terms, $\delta_{c,s}$, $\delta_{c,\nu}$, and $\delta_{c,j}$ (inner dark green shading, intermediate light green shading, and external dark green shading, respectively). The two symbols with 1$\sigma$ error bars labeled ``GCLASS''
represent the results the dynamical analysis of GCLASS clusters, as obtained by \citetalias{Biviano+16} (diamond) and as obtained in this work by including the new GOGREEN survey data for the GCLASS clusters. The symbols and 1$\sigma$ error bars labelled ``Lensing'' and ``X-ray'' represent mean values from the literature for clusters in the same $\rvir$ and $z$ range as the clusters in our sample, obtained through the analysis of gravitational lensing and X-ray data, respectively
(references are given in the text). {\em Bottom panel:} Green horizontal line and shadings as in the top panel. Symbols with 1$\sigma$ error bars are theoretical predictions from \citet{BHHV13,DM14,CWSD15,Child+18,DeBoni+13,Ragagnin+21} 
for a halo of the same $\mvir$ and $z$ as our {\em ensemble} cluster, and rescaled to the cosmological parameters used in this paper.}
\label{f:cwhole}
\end{figure}

In Fig.~\ref{f:cwhole} we show the $\cal{L}$-weighted average $\cvir$ for the {\em ensemble} cluster. In the top panel of the same figure we also display three other observational values of $\cvir$ from the literature  (gray diamonds). One of them is the average value estimated by \citetalias{Biviano+16} for the GCLASS sample. Next to this value we also plot our new determination of $\cvir$ for the same GCLASS sample (gray dot), including new GOGREEN survey data from the GCLASS clusters. Our new determination is very close to the original one by \citetalias{Biviano+16}.

The other two literature values shown in Fig.~\ref{f:cwhole} are obtained by considering the 20 clusters with gravitational lensing determinations of $\cvir$ and the 8 clusters with X-ray data based determinations of $\cvir$, within the $z$ and $\rvir$ ranges of our sample. 
We refer to these two literature values as ``Lensing'' and ``X-ray,'' respectively.
When a cluster has more than one $\cvir$ determination listed, we assign the cluster the error-weighted average of its $\cvir$ values. We then compute the weighted average of all the literature values and its error, using the inverse of the listed $\cvir$ errors as weights. 
The $\cvir$ values on which these two averages are computed come from \citet{AECS16}, \citet{BDPV14}, \citet{Jee+06}, \citet{Jee+11}, \citet{MLWS05}, and \citet{SGEM15} and the compilations of \citet{SC13} and \citet{GGS16}.

In the bottom panel of Fig.~\ref{f:cwhole} we also show the theoretical predictions for $\cvir$ of a halo with the same $\mvir$ and $z$ as our {\em ensemble} cluster, as
obtained from numerical simulations. We consider, in particular, the theoretical predictions of \citet{DeBoni+13} and \citet{Ragagnin+21}, which are based on hydrodynamical simulations, and those of \citet{BHHV13,DM14,CWSD15,Child+18}, which  are based on gravity-only simulations. Since these simulations adopted different cosmological parameters, we use eq.~(7) and Table~2 in \citet{Ragagnin+21} to convert the theoretical $\cvir$ values to those expected in the assumed cosmology in this paper. The value of $\sigma_8$ does not enter in our dynamical analysis; we assume $\sigma_8=0.8$. We neglect the effect of a change in the baryonic mass density, as this is sub-dominant with respect to changes in $\Omega_m$ and $\sigma_8$. After transformation to the same cosmological parameters, the modifications to the published theoretical $\cvir$ values range from $-2$ to $+11$~\% and the rms among the different values decreases by 10\%, from 0.36 to 0.33. The remaining differences can result from the inclusion or not of baryonic physics in the simulations \citep{DeBoni+13} and/or from the way the parameter $\cvir$ has been measured \citep{DeBoni+13,Child+18}. According to \citet{BHHV13} and \citet{Child+18} the distribution of $\cvir$ among cosmological halos has a dispersion of $\simeq \cvir/3$, so the expected theoretical uncertainty for a $\cvir$ value based on a sample of $N_c$ halos is $\cvir / (3 \, \sqrt{N_c})$, which defines the error bars on the theoretical values shown in Fig.~\ref{f:cwhole}. 

The comparison with previous observational determinations show that our new determination of $\cvir$ is in good agreement with that of \citetalias{Biviano+16}, only slightly lower, and  more precise, thanks to our larger data set. This agreement is not surprising, given that the sample of GCLASS clusters investigated by \citetalias{Biviano+16} is included in our sample. Our $\cvir$ value is also in good agreement with the Lensing and X-ray literature values, and it lies between the two estimates.

Our new determination of $\cvir$ is also in excellent agreement with most theoretical predictions, in particular with the most recent ones, based on either DM-only or on hydrodynamical simulations. Theoretically predicted $\cvir$ values of cluster-size halos have a very mild dependence on baryonic processes.

\begin{figure}
\centering
\includegraphics[width=.5\textwidth]{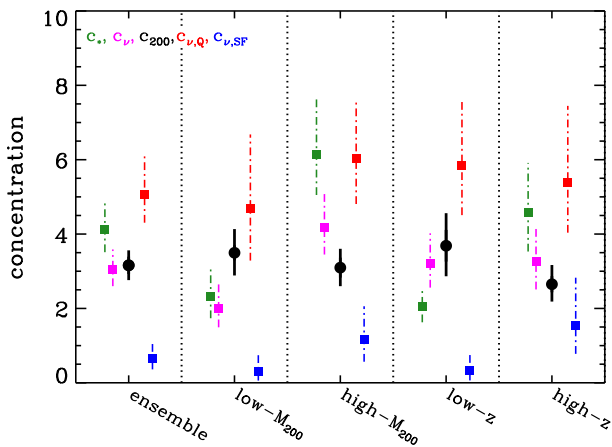} 
\caption{Concentrations  ($\cvir$) for the {\em ensemble} cluster and its subsamples (black dots with 1$\sigma$ error bars).
The green, magenta, red, and blue squares with dot-dashed 1$\sigma$ error bars represent $c_{\star}, c_{\nu}, c_{\nu,\rm{Q}}$, and $c_{\nu,\rm{SF}}$, respectively.}
\label{f:concs}
\end{figure}

We do not find any significant difference between the $\cvir$ values of the {\em low-}$z$ and {\em high-}$z$ subsamples, nor between the $\cvir$ values of the {\em low-}$\mvir$ and {\em high-}$\mvir$ subsamples (see Table~\ref{t:concs} and Fig.~\ref{f:concs}). This is in line with theoretical expectations \citep{Ragagnin+21}, according to which $\cvir$ changes by $<0.2$ in the $z$ and $\mvir$ ranges of our four subsamples (see Table~\ref{t:ensemble}, Cols. 5 and 7); that is, the predicted changes are less than our observational uncertainties.

We then compare the $\cvir$ values found for the different subsamples with the concentrations obtained for the number density profiles of all, quiescent, and star-forming galaxies, $c_{\nu}, c_{\nu,\rm{Q}}, c_{\nu,\rm{SF}}$, respectively, and  with the concentration obtained for the stellar mass density profiles using all galaxies, $c_{\star}$ (see Table~\ref{t:concs} and Fig.~\ref{f:concs}). The values of $\cvir$ are consistent within 2$\sigma$ with the values of $c_{\nu}$ and also with the values of $c_{\nu,\rm{Q}}$, for all (sub)samples. However, $\cvir > c_{\nu,\rm{SF}}$ at $>3 \sigma$ level for the {\em ensemble} cluster and the {\em low-}$M_{200}$ and {\em low-}$z$ subsamples. Cluster galaxies therefore seem, in general, to trace the mass distribution if they are not star-forming.

Previous results for clusters at lower-$z$ have found only a mild radial dependence of the total mass per galaxy in clusters \citep[e.g.,][]{BG03,LMS04,Annunziatella+14}, which is not in conflict with our result, given the large uncertainties in our concentration values. \citetalias{Biviano+16} found a higher stellar mass concentration compared to the total mass concentration, but we do not confirm their finding.

We do not find any significant difference between $c_{\nu}$ and $c_{\star}$ in any of our subsamples, in agreement with \citet{vanderBurg+14} who did not find a significant difference between $c_{\nu}$ and $c_{\star}$ in GCLASS clusters at $z \sim 1$. For the {\em ensemble} cluster and its {\em low-}$z$ and {\em high-}$M_{200}$ subsamples, $c_{\nu,\rm{Q}} > c_{\nu,\rm{SF}}$ at $> 3 \sigma$. Quiescent galaxies tend to have a more concentrated spatial distribution than star-forming galaxies, as expected from previous works \citep[e.g.,][]{Gisler78,Biviano+97,Dressler+13,vanderBurg+14}. This difference is not found to be statistically significant in all of our subsamples, likely due to the large uncertainties. 

\section{The velocity anisotropy profiles}\label{s:beta}
\subsection{Method: The Jeans inversion technique}\label{ss:jeansinvmeth}
With MAMPOSSt it is possible to constrain $\mr$ and $\br$ at the same time, but only within the limits allowed by the restricted set of models chosen. The choice of $\mr$ models is facilitated by the large number of works that have determined cluster and halo $\mr$ from observations and cosmological simulations \citep[e.g.,][and references therein]{Ludlow+13,Pratt+19}. On the other hand, simulations suggest that there is a large variance in the shape of $\br$ among different cluster-size halos \citep[see Fig.~1 in][]{MBB13}. There is no guarantee that the four $\br$ models we adopt in the MAMPOSSt analysis  (see Sect.~\ref{ss:MAMmeth}) are a close fit to the real, average cluster $\br$.

We can go beyond the adopted $\br$ models by performing the inversion of the Jeans equation by the method of \citet{BM82} in the implementations of \citet{SSS90} and \citet{DM92}. The inversion of the Jeans equation requires knowledge of $\mr$. We adopt the $\cal{L}$-weighted average of all 35 MAMPOSSt best-fit model $\mr$. The inversion procedure also requires knowledge of two observables: the projected number density profile $N(R)$ and the line-of-sight velocity dispersion profile $\sigma_{\rm{los}}(R)$. In practice, some smoothing of the observed profiles is needed to run the Jeans inversion algorithm, and we use the LOWESS smoothing technique \citep[see, e.g.,][]{Gebhardt+94}. The number density profile is then deprojected numerically \citep[using Abel's equation; see][]{BT87}. Since the equations to be solved contain integrals up to infinity, we extrapolate the profiles with the functional form of eq.~(10) in \citet{Biviano+13}, out to 30 Mpc; we  confirmed that  this radius is large enough for the results to be stable.

Uncertainties in the $\beta(r)$ are estimated by performing the Jeans inversion on 100 bootstrap resamplings of the original data sets. The average of the 100 bootstrap resamplings is finally taken as our fiducial $\beta(r)$ to correct for possible bias in the original estimate \citep{ET86}. The resulting $\br$ is finally compared with the $\cal{L}$-weighted average of all 35 MAMPOSSt best-fit model $\br$ to check for consistency in the procedure.

\subsection{Results}\label{ss:jeansinvres}
\begin{figure}
\centering
\includegraphics[width=.5\textwidth]{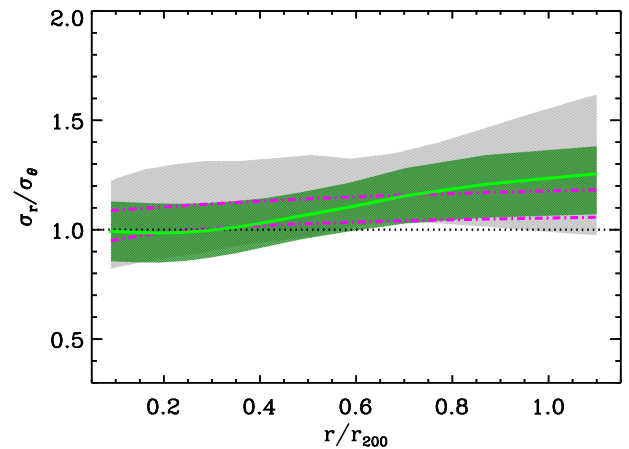} 
\caption{Velocity anisotropy profile $(\bpr)(r)$ for all galaxies of the {\em ensemble} cluster (pale green line) and 1$\sigma$ confidence level (dark green shading). Dot-dashed magenta lines: 1$\sigma$ intervals of the ${\cal L}$-weighted average of the 35 MAMPOSSt best-fit model $(\bpr)(r)$. Gray shading: 1$\sigma$ intervals of the result of \citetalias{Biviano+16} for the GCLASS sample.}
\label{f:betawhole}
\end{figure}

We show in Fig.~\ref{f:betawhole} the velocity anisotropy profile for all galaxies in the {\em ensemble} cluster (green line and shading). Rather than displaying $\br$, we display $(\bpr)(r)$ since this represents the relative importance of the radial and tangential components of the velocity dispersion in a more linear way. Figure~\ref{f:betawhole} shows that $(\bpr)(r)$ increases from $\simeq 1$ near the cluster center to slightly $>1$ at $\rvir$, which means that  the orbits of cluster members are isotropic near the cluster center and become more radially elongated at larger radii. The Jeans inversion result is fully consistent with the MAMPOSSt result (magenta dash-dotted curves). In Fig.~\ref{f:betawhole} we also compare our new result with that found for the GCLASS sample by \citetalias{Biviano+16} (gray shading). We find full consistency between the two profiles, the new one being more precise thanks to the increased statistics, now showing stronger evidence for non-isotropic orbits at large radii.

Our new $\br$ determination is the highest-$z$ so far obtained for galaxies in clusters. It is very similar to other $\br$ found for galaxies in lower-$z$ clusters \citep[e.g.,][]{NK96,BK03,BP09,Lemze+12,Biviano+13,Annunziatella+16,Capasso+19,Mamon+19}.

\begin{figure}
\centering
\includegraphics[width=.5\textwidth]{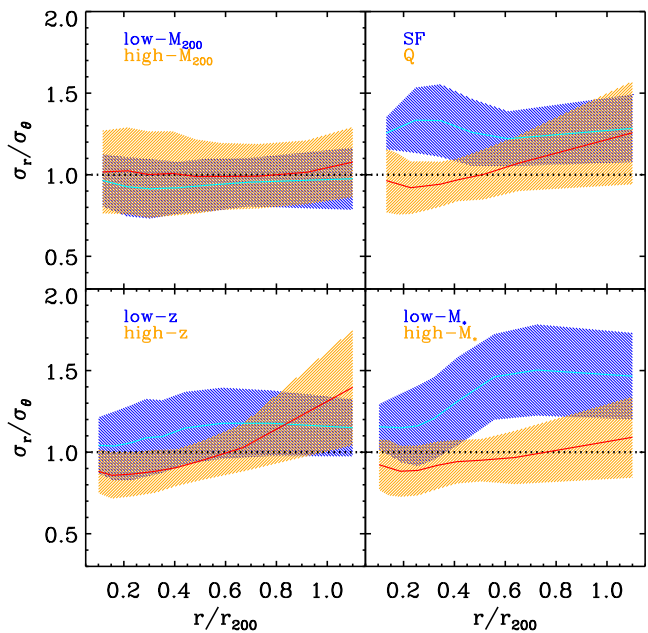} 
\caption{Velocity anisotropy profiles $(\bpr)(r)$ for several subsamples. Confidence intervals are 1$\sigma$.
{\it Upper left panel:} {\em low-}$\mvir$ (cyan line, blue shading) and {\em high-}$\mvir$ (red line, orange shading). {\it Lower left panel:} {\em low-}$z$ (cyan line, blue shading) and {\em high-}$z$ (red line, orange shading). {\it Upper right panel:} star-forming galaxies (cyan line, blue shading) and quiescent galaxies (red line, orange shading) in the {\em ensemble} sample. {\it Lower right panel:} $9.5 \leq \log M_{\star}/{\rm M}_{\odot} < 10.5$ galaxies (cyan line, blue shading) and $\log M_{\star}/{\rm M}_{\odot} \geq 10.5$
galaxies (red line, orange shading) in the {\em ensemble} sample.}
\label{f:betasubsam}
\end{figure} 

We also determine the $\br$ of eight subsamples of the {\em ensemble} cluster (see Fig.~\ref{f:betasubsam}). Splitting the {\em ensemble} cluster into its subsamples reduces the statistics and increases the error bars on the profiles, so in general  it is    impossible to find significant differences among the subsamples. There is no significant difference between the orbits of galaxies in {\em low-}$\mvir$ and {\em high-}$\mvir$ clusters (see  upper left panel of Fig.~\ref{f:betasubsam}) nor between the orbits of galaxies in {\em low}-$z$ and {\em high-}$z$ clusters (see lower left panel of Fig.~\ref{f:betasubsam}). These results are in agreement with those of \citet{Capasso+19}. Theoretical expectations are for more radial orbits in more massive and higher-$z$ cluster-sized halos \citep[see Fig.~10 in][]{Munari+13}; we might be unable to distinguish any such trend because of the limited statistics and the limited $z$- and $\mvir$-range of our clusters.

We find that star-forming galaxies have more radially elongated orbits than quiescent galaxies (see upper right panel of Fig.~\ref{f:betasubsam}), in agreement with previous results \citep{BK03,BP09,Biviano+13,MBM14,Mamon+19}, but this difference is only marginally significant. Slightly more significant is the result that low-$M_{\star}$ ($9.5 \leq \log M_{\star}/{\rm M}_{\odot} < 10.5$) galaxies have more radially elongated orbits than high-$M_{\star}$ ($\log M_{\star}/{\rm M}_{\odot} \geq 10.5$) galaxies (see lower right panel of Fig.~\ref{f:betasubsam}). Given that star-forming galaxies are, on average, less massive than quiescent ones, the two  results given above are not independent. Taken at face value, the apparent difference we find between low- and high-$M_{\star}$ galaxy orbits is in the opposite sense of that found by \citet{Annunziatella+16} in a $z=0.2$ cluster. However, the sample of \citet{Annunziatella+16} only considered quiescent galaxies (defined based on galaxy colors) and had a lower cut in $M_{\star}$, so a direct comparison with the current result is difficult and perhaps not very meaningful.

The orbital difference between quiescent and star-forming galaxies has generally been interpreted as being the consequence of a different time of accretion into the cluster of the two populations, with the star-forming, less-massive galaxies being recent infallers \citep{Lotz+19}. Those that we observe today as massive quiescent galaxies might have entered the cluster at an earlier phase of its evolution when collective collisions were able to isotropize galaxy orbits \citep{LC11}. The lower-velocity dispersion of clusters at an earlier phase might have been more suitable to the occurrence of galaxy-galaxy mergers, which could have played a role in dissipating part of the orbital energy of these galaxies.

\section{The pseudo phase-space density profiles}\label{s:PPSD}
By combining the results of Sect.~\ref{ss:MAMres} on $M(r)$ and Sect.~\ref{ss:jeansinvres} on $\beta(r)$, in this section we  determine the pseudo phase-space density profiles $Q(r)$ and $\qr(r)$. These are obtained as the ratio of either the total mass density or the number density profile to the cube of either the total (for $Q(r)$) or the radial (for $\qr(r)$) velocity dispersion profile. 

\begin{figure}
\centering
\includegraphics[width=.5\textwidth]{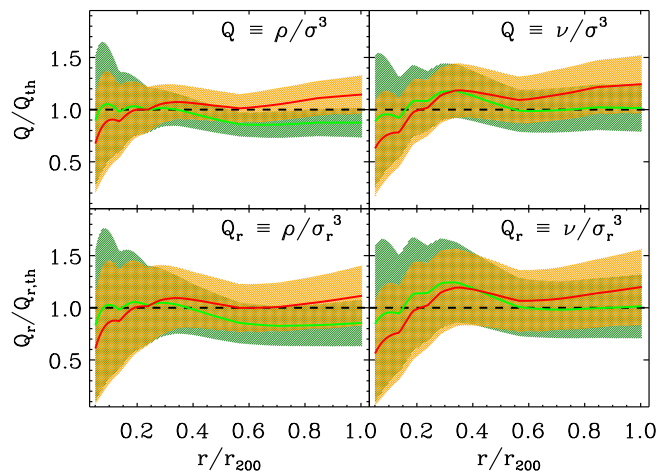} 
\caption{The pseudo phases-space density profiles of the {\em ensemble} cluster.
  The solid lines show $Q(r)$ and $\qr(r)$ of the {\em ensemble} cluster (top and bottom panels, respectively)  derived from the total mass density profile $\rho(r)$ and the number density profile $\nu(r)$ (left and right panels, respectively), divided by power laws of the radius $\qt \propto r^{-1.84}$ and $\qtr \propto r^{-1.92}$ (green line and shadings) or by power laws $\qt \propto r^{-2.02}$ and $\qtr \propto r^{-2.11}$ (red line and orange shading). The shadings indicate 1$\sigma$ uncertainties. The dashed black line is unity.}
\label{f:ppsdp}
\end{figure} 

We determine the total mass density profile $\rho(r)$ from differentiation of the ${\cal L}$-weighted average of the 35 MAMPOSSt best-fit $M(r)$ (see Sect.~\ref{ss:MAMres}), and obtain its 1$\sigma$ uncertainty from the ${\cal L}$-weighted dispersion of the corresponding 35 $\rho(r)$. 
We determine the number density profile $\nu(r)$ by Abel inversion of the projected number density profile $N(R)$ (see Sect.~\ref{ss:jeansinvmeth}). We assume for $\nu(r)$ the same fractional uncertainties of $N(R)$. Within the Jeans inversion procedure (see Sect.~\ref{ss:jeansinvmeth}) we not only determine $\beta(r),$ but also the total 3D velocity dispersion profile $\sigma(r)$ and radial velocity dispersion profile $\sigma_{{\rm r}}(r)$ \citep[see eqs. 20 and 21 in][]{SSS90}, and their uncertainties via the already described bootstrap technique.

In Fig.~\ref{f:ppsdp} we show the {\em ensemble} cluster $Q(r)$ and $\qr(r)$  divided by power laws of the radius $\qt \propto r^{-\alpha}$ and $\qtr \propto r^{-\ar}$, respectively. We show in green the results for $\alpha=1.84$ and $\ar=1.92$, the values found by \citet{DML05} based on the cosmological DM-only simulations of \citet{DMS04_rho,DMS04_vel}. According to \citet[][see their Fig.~6]{LC09} these values evolve with $z$, increasing by a factor $\simeq 1.1$ at the mean redshift of our sample. The results for these evolved values of $\alpha$ and $\ar$ are shown in red in Fig.~\ref{f:ppsdp}. The power laws have been normalized to the median value of $Q(r)$ and $\qr$ over the radial range 0.05--1 $\rvir$. We show the results for $\rho(r)$ in the left panels and those for $\nu(r)$ in the right panels. 

We see from Fig.~\ref{f:ppsdp} that the observational profiles do not deviate significantly from the theoretical power laws. Given the current observational uncertainties we are unable to claim which of the considered theoretical slopes is more consistent with our data. 

\section{Discussion}\label{s:disc}
Our dynamical analysis indicates that the velocity distribution of the members of our $z \simeq 1.1$ {\em ensemble} cluster (Fig.~\ref{f:vdpmamp}) can be nicely described by a spherical system of collisionless tracers in dynamical equilibrium on mildly radial anisotropic orbits (Fig.~\ref{f:betawhole}) in a gravitational potential that is quite close to theoretical predictions for cluster-size halos in a $\Lambda$CDM cosmology (Figs.~\ref{f:mr} and \ref{f:cwhole}). 

Our results have been obtained for an {\em ensemble} cluster built by stacking 14 different clusters. Even if theory predicts that clusters   form a nearly homologous set modulated mostly by $r_{\Delta}$ and only mildly by $c_{\Delta}$, there is observational evidence that suggests some variance in the dynamical structure of clusters of similar mass and redshift \citep{Donahue+02,Biviano+17b,Lopes+18}. Simulations indicate that the scatter in the cluster mass density profiles has a Gaussian distribution, so the technique of stacking halos should not suffer from bias \citep{RKG11}. Additional support for the use of a stack sample comes from the analysis of simulated halos by the MAMPOSSt technique. \citetalias{MBB13} have shown that the recovered $\cvir$ from a stack of 11 simulated clusters does not suffer from a statistically significant bias.
To address the importance of the variance in the cluster mass density profiles, we  ran a jackknife analysis. We found that sample variance does not dominate the error budget on the derived $\cvir$ (see Sect.~\ref{ss:MAMres} and Table~\ref{t:concs}). However, our results for the {\em ensemble} cluster must be considered valid only in an average sense. 

\subsection{The mass profile}\label{ss:discmr}
The {\em ensemble} cluster $M(r)$ is very close to the NFW model (Fig.~\ref{f:mr}). Although the Burkert (core) model is preferred over the NFW model by the MAMPOSSt analysis, the difference between the two models is not significant, and   other models (Einasto, Hernquist, gNFW) are in fact equally acceptable. It therefore looks like NFW is an acceptable description of the mass distribution in clusters of galaxies from $z \sim 0$ \citep[e.g.,][]{Carlberg+97-mprof,KBM04} to $z \simeq 1.1$, although other models could be acceptable as well.

The {\em ensemble} cluster $\cvir$ is in very good agreement with, albeit slightly smaller than, the corresponding value obtained by \citetalias{Biviano+16} for the GCLASS clusters by a similar analysis. Our new $\cvir$ value is also in good agreement with previous estimates for clusters in the same $z$ and $\rvir$ range of our 14 clusters, and based on gravitational lensing and X-ray data. It has been argued that $\cvir$ estimated from gravitational lensing \citep{Meneghetti+11,Meneghetti+14} and X-ray \citep{RBEMM13} data might be biased high because of selection effects. More concentrated clusters, and clusters with their major axis aligned along the line of sight, are stronger gravitational lenses and have higher X-ray luminosity per given mass. The similarity of the $\cvir$ values obtained via three different methods suggests that the current sample of $z \gtrsim 1$ clusters for which $\cvir$ have been measured is representative of the whole population in the given $\rvir$ range.

Our new $\cvir$ value is in excellent agreement with several predictions from cosmological simulations for a halo with the same mass and redshift as our {\em ensemble} cluster (see Fig.~\ref{f:cwhole}, Table~\ref{t:ensemble} and Table~\ref{t:concs}). The agreement is particularly good with the most recent theoretical predictions, based on either DM-only or hydrodynamical simulations. We note that there is very little difference among the $\cvir$ values predicted by DM-only \citep{Child+18} or by hydrodynamical simulations, indipendently of whether the latter account \citep{Ragagnin+21} or not \citep{DeBoni+13} for feedback by an active galactic nucleus (AGN)  in the central cluster region. 
The addition of baryonic effects to DM-only simulations can, in principle, lead to a substantial change in the halo $\cvir$ \citep{Fedeli12,Cui+16}. However, adiabatic contraction and stellar and AGN feedback have the opposite effect on $\cvir$ \citep{TMMDM11}, so the difference between a cluster-size halo $\cvir$ in DM-only
and full hydrodynamical simulations turns out to be quite small
\citet{KM11,RBEMM13}.

Given the current observational uncertainties on $\cvir$ ($\sim 20$\%; see Table~\ref{t:concs}) we cannot use it to constrain the relative importance of different feedback schemes. 
Since the effects of AGN feedback are strongest near the cluster center \citep[e.g.,][]{MTM13,Schaller+15,Peirani+17,SLN18}, better constraints can be obtained by the determination of the inner slope of the total mass density profile. As already pointed out, our {\em ensemble} cluster $M(r)$ is better represented by a cored than a cuspy model (Fig.~\ref{f:mr}). Taken at face value, this result would suggest that AGN feedback is required. However, we cannot make the difference between a cored and a cuspy $M(r)$ in a statistically significant way because we lack information on the kinematics in the inner $\sim 50$ kpc that could be provided by observation of the stellar velocity distribution of the BCG \citep{Sand+04,Newman+13,Sartoris+20}.

The concentration of the total mass profile of our {\em ensemble} cluster $\cvir$ is not significantly different from either the concentration of the number density profile $c_{\nu}$ or from the concentration of the $M_{\star}$ density profile $c_{\star}$.
\citetalias{Biviano+16}  claimed $\cvir < c_{\star}$ for the sample of GCLASS clusters and attributed this to an excess of massive galaxies near the center of $z \sim 1$ clusters.
Our {\em low-}$z$ subsample has $z=0.96$, similar to that of the GCLASS sample analyzed by \citetalias{Biviano+16}, yet we do not see any difference between $\cvir$ and $c_{\star}$ in this subsample either. Therefore, the difference with respect to \citetalias{Biviano+16} cannot be attributed to a difference in $z$ between our sample and theirs.  However, the significance of the claimed $\cvir$ versus $c_{\star}$ difference in \citetalias{Biviano+16} is only $\lesssim 2 \sigma$, and our values of $\cvir$ and $c_{\star}$  are consistent with the respective values of \citetalias{Biviano+16} within $\leq 1.3 \sigma$. So we conclude that there is no real tension between our results and those of \citetalias{Biviano+16}, in terms of relative total and stellar mass concentrations. 

The similarity of the number density, stellar mass, and total mass concentrations can be rephrased by saying that the distribution of cluster galaxies appears to follow that of the total mass (or vice versa). Lower-$z$ cluster studies have found only a mild difference in the two distributions \citep[e.g.,][]{BG03,LMS04,Annunziatella+14}, so our result suggests that galaxies retain a similar distribution to that of the mass from $z \sim 0$ to $z \simeq 1.1$. This similarity is broken when only star-forming galaxies are considered. We find $c_{\nu,\rm{Q}} > c_{\nu,\rm{SF}}$, as also seen in lower-$z$ clusters \citep[e.g.,][]{Gisler78,Biviano+97,Dressler+13,vanderBurg+14}, an indication that the spatial segregation of quiescent and star-forming galaxies is already in place at $z \gtrsim 1$.

\subsection{The velocity anisotropy profile}\label{ss:discbr}
The {\em ensemble} cluster $\beta(r)$ is fully consistent (but with smaller error bars thanks to the better statistics) with the result obtained by \citetalias{Biviano+16} for the GCLASS sample. Orbits are isotropic near the center, and $\br>0$ ($\bpr>1$) at $r>0.4 \, \rvir$. The evidence for non-isotropic, radial orbits, at $r \simeq \rvir$ is partially significant. 

Our {\em ensemble} cluster $\br$ is very similar to that of DM particles in cluster-size cosmological halos \citep[see Fig.~\ref{f:betawhole} and][]{MBM10,Munari+13}. The central isotropy ($\br \simeq 0$ or $\bpr \simeq 1$) is predicted to be the result of collective collisions taking place in the initial phases of the cluster formation, with the radial extent of the isotropic region being dependent on the efficiency of these collective collisions \citep{LC11}. At large radii, where cluster growth is mostly governed by smooth accretion, the radial accretion of material is reflected in $\br > 0$ \citep{LC11}.

Different subsamples of the {\em ensemble} cluster do not show significantly different $\br$ (see left panels of Fig.~\ref{f:betasubsam}). On the other hand, different classes of cluster galaxies do show partially significant different $\br$ (see right panels of Fig.~\ref{f:betasubsam}). The orbits of star-forming  and low-$M_{\star}$ ($9.5 <\log M_{\star}/{\rm M}_{\odot} <10.5$) galaxies are more radially elongated than those of quiescent and high-$M_{\star}$ ($\log M_{\star}/{\rm M}_{\odot} \geq 10.5$) galaxies. A similar difference is seen in most \citep[but not all; see][]{AADDV17} lower-$z$ clusters considering galaxies of different morphological \citep{Mamon+19} and spectral type \citep{BK03,BP09,Biviano+13}, and different colors \citep{MBM14}. 

A possible interpretation of the orbital difference between quiescent and star-forming galaxies relies on a different time of accretion into the cluster of the two populations, star-forming less-massive galaxies being recent infallers \citep{Lotz+19}. At an earlier time the cluster gravitational potential was rapidly changing, and the cluster velocity dispersion was sufficiently small to make galaxy-galaxy mergers possible. So galaxies residing in clusters for a long time could have had their orbits isotropized by collective collisions and mergers \citep{LC11}. These galaxies would then evolve into massive  quiescent galaxies.

\subsection{The pseudo phase-space density profile}\label{ss:discqr}
The pseudo-phase space density profiles, $Q(r)$ and $\qr(r)$, of the {\em ensemble} cluster are fully consistent with the predicted power-law behavior for cluster-sized halos \citep[][see Fig.~\ref{f:ppsdp}]{TN01,RTM04,DML05,LC09}, both when using $\rho(r)$ and when using $\nu(r)$ in the definition of $Q(r)$ and $\qr(r)$. Our result confirms the previous results obtained for clusters in the range $0 \lesssim z \lesssim 0.9$ \citep{Biviano+13,MBM14,AADDV17,Capasso+19} and for the $z \sim 1$ GCLASS cluster sample \citepalias{Biviano+16}. 

The scale-free behavior of $Q \propto r^{\alpha}$ and $\qr \propto r^{\ar}$, and even the value of $\alpha$, were predicted by the \citet{Bertschinger85} self-similar solution for secondary infall of material onto a point-mass perturber in a uniformly expanding universe. Cluster assembly is  thought to proceed in a more chaotic way, however, so other explanations are needed for the power-law behavior of $Q(r)$ and $\qr(r)$. According to \citet{DML05}, the scale-free behavior of $\qr(r)$ emerges from an initial scale-free phase-space density that keeps its scale-free behavior after violent relaxation because this process is driven by gravity alone. The system evolves to dynamical equilibrium (as described by the Jeans equation) $\ar=1.94 + 0.22 \, \beta(0)$, thus for $\beta(0) \simeq 0$, which is what we find (see Fig.~\ref{f:betawhole}), $\ar$ is very close to the value we observe in our {\em ensemble} cluster (see Fig.~\ref{f:ppsdp}).

\begin{figure}
\centering
\includegraphics[width=.5\textwidth]{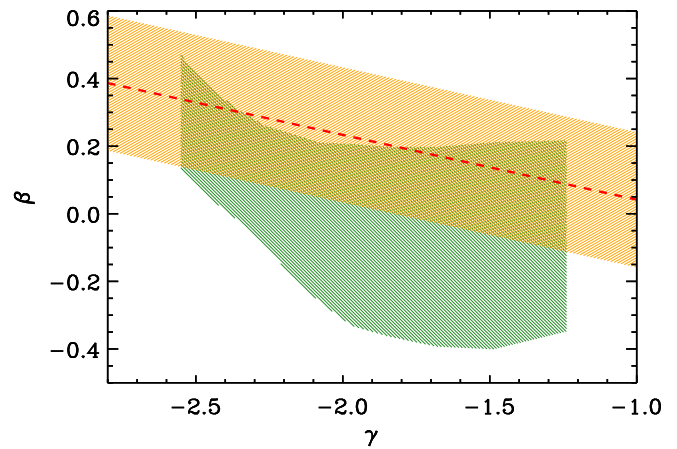} 
\caption{$\beta-\gamma$ relation of \citet{HM06} with its $\pm 1 \sigma$ confidence interval (dashed red line and orange shading). The green shading indicates the $\pm 1 \sigma$ confidence interval for our {\em ensemble} cluster.}
\label{f:betagamma}
\end{figure} 

The analytical solution found by \citet{DML05} for $\qr(r)$ is based on the assumption of a linear relation between $\gamma(r)$ (the logarithmic slope of $\rho(r)$) and $\beta(r)$,
as indicated by the analysis of simulated halos \citep{HM06}. In Fig.~\ref{f:betagamma} we show that $\br$ and $\gamma(r)$ of our {\em ensemble} cluster are indeed consistent with a linear relation, although slightly below that
predicted by \citet{HM06}\footnote{It has been argued by \citet{AW20} that the scale-free behavior of $Q(r)$ is a mere coincidence that results from the Einasto shape of $M(r)$ (we do find that the Einasto model is a good representation of our {\em ensemble} cluster $M(r)$; see Sect.~\ref{ss:MAMres}) and the linear $\beta-\gamma$ relation. However, we do not see why the Einasto model or the linear $\beta-\gamma$ relation should be considered more fundamental than the scale-free behavior of $Q(r)$.}.

On average, our results for $M(r), \beta(r), Q(r),$ and $\qr(r)$ indicate that the clusters in our sample have already reached dynamical equilibrium. The clusters in our sample are not representative of the whole cluster population at $z \simeq 1.1$ since they are selected based on the presence of a red-sequence galaxy population (the SpARCS clusters), and a hot intra-cluster gas (the SPT clusters), and both characteristics are typical of mature galaxy systems. In this respect we conclude that at least those
$z \simeq 1.1$ clusters that contain a hot intra-cluster medium and an evolved galaxy population are also dynamically relaxed.

The highest-$z$ clusters known are at $z \lesssim 2$ \citep{Stanford+12,Andreon+14} and the $z>2$ regime is generally considered to be the realm of proto-clusters \citep{Overzier16}. In the hypothesis that, on average, massive clusters start assembling at $z \sim 2$, our results indicate that the $\sim 2$ Gyr  between $z \sim 2$ and $z \sim 1$ is sufficient time for clusters to reach dynamical equilibrium after initial assembly. This time is similar to the dynamical time at $z=1.1$, $t_{{\rm dyn}} \simeq (G \Delta \rho_c(z))^{-1/2} = 1.5$ Gyr \citep{Sarazin86}, for an overdensity $\Delta=200$, where
$\rho_c(z)$ is the critical density at redshift $z$. The origin of the universal shape of $M(r), Q(r)$, and $\qr$ could therefore result from a process that ensures fast dynamical relaxation of clusters at their formation, such as violent relaxation and chaotic mixing \citep{Henon64,LyndenBell67,LC11,BdSV19}. The orbital shape of cluster galaxies also seems to be largely imprinted after this violent relaxation process since the $\beta(r)$ we find for our {\em ensemble} cluster is quite similar to those found for lower-$z$ clusters, at least down to $z \simeq 0.2$ \citep{Annunziatella+16,Capasso+19}. 

Our finding that $z \simeq 1.1$ clusters are dynamically old appears to fit well in the scenario supported by the analyses of the cluster galaxy population based on GOGREEN survey data. These studies \citep{Old+20,Old+21,vanderBurg+20,Webb+20} have in fact shown that the GOGREEN clusters are also old in terms of the age of their member galaxies. \\

\section{Conclusions}\label{s:conc}
We investigated the internal dynamics of 14 clusters at $0.9 \leq z \leq 1.4$ drawn from the GOGREEN spectroscopic data set \citep{Balogh+17,Balogh+21} complemented with data from the GCLASS survey \citep{Muzzin+12}. We stacked the 14 clusters to build an {\em ensemble} cluster containing 581 member galaxies with $M_{\star} \geq 10^{9.5} {\rm M}_{\odot}$. We used MAMPOSSt \citepalias{MBB13} to determine the {\em ensemble} cluster mass profile $M(r)$, and we inverted the Jeans equation with the method of \citet{SSS90} to determine the {\em ensemble} cluster velocity anisotropy profile $\beta(r)$. Using the results of the MAMPOSSt and Jeans inversion analyses, we determined the pseudo phase-space density profiles $Q(r)$ and $\qr(r)$. We also considered four subsamples of the {\em ensemble} cluster by separating the 14 clusters that compose it into two groups of two, split by $\mvir$ or by $z$. Our results are the following:
\begin{itemize}
\item Several $M(r)$ models are acceptable and we cannot discriminate between cored and cuspy models, presumably because we lack dynamical tracers in the very inner region ($< 0.05$ Mpc) that could be provided by observations of the BCG stellar kinematics \citep[as in][]{Sartoris+20}.
\item The concentration $\cvir$ of the {\em ensemble} cluster mass profile is close to theoretical predictions, and to previous observational determinations from the literature for clusters at similar redshifts, obtained using gravitational lensing and X-ray data. 
\item The value of $\cvir$ does not depend on the cluster redshift or mass, as expected from theory, which predicts a very small change in $\cvir$ over the redshift and mass ranges spanned by our cluster sample.
\item The total mass concentration is not significantly different from the concentration of the spatial distribution of galaxies, or from the stellar mass concentration.
\item The star-forming galaxies have a less concentrated distribution than the quiescent galaxies.
\item The orbits of cluster galaxies are isotropic near the center and more radial outside. \item High-$M_{\star}$ and quiescent galaxies have more isotropic orbits than, respectively, low-$M_{\star}$ and star-forming galaxies, but this orbital difference is only marginally significant.
\item The profiles $Q(r)$ and $\qr(r)$, determined either using the total mass or the number density profile, are very close to theoretical power-law predictions. 
\end{itemize}

We conclude that the internal dynamics of clusters at the highest redshift probed so far in detail, do not differ from the internal dynamics of lower-redshift clusters, and confirm theoretical predictions. 
The fundamental dynamical properties of clusters, such as the shape of their mass density profile, the orbits of their galaxies, and the power-law behavior of the pseudo phase-space density profile have been in place since $z \lesssim 1.4$. Given that very few clusters are known to exist beyond $z \sim 2$, clusters must reach their dynamical equilibrium configuration in $\sim 2$ Gyr, presumably by some rapid process of dynamical relaxation \citep{LyndenBell67,LC11}. 

Our work is based on a limited sample of 14 clusters, selected because of the presence of a red sequence of cluster members, or of a hot intra-cluster medium. The similarity of our $\cvir$ value to those determined using different methods suggests that we are not dealing with a biased cluster sample selection. Nevertheless, we caution that our conclusions might not be applicable to the whole population of $z \simeq 1.1$ clusters, and future investigations are needed to confirm our results with a statistically representative sample. 

\begin{acknowledgements}
This work was enabled by observations made from the Gemini North, Subaru and CFHT telescopes, located within the Maunakea Science Reserve and adjacent to the summit of Maunakea. We are grateful for the privilege of observing the Universe from a place that is unique in both its astronomical quality and its cultural significance. We thank the referee for her/his useful comments that helped improving the scientific content of this paper. 
RD gratefully acknowledges support from the Chilean Centro de Excelencia en Astrof\'isica y Tecnolog\'ias Afines (CATA) BASAL grant AFB-170002. JN received support from Universidad Andr\'es Bello research grant DI-12-19/R. LJO acknowledges the support of a European Space Agency (ESA) Research Fellowship. GR acknowledges support from the National Science Foundation grants AST-1517815, AST-1716690,and AST-1814159 and NASA HST grant AR-14310. GR also acknowledges the support of an ESO visiting science fellowship.  BV acknowledges financial contribution from  the grant PRIN MIUR 2017 n.20173ML3WW\_001 (PI Cimatti) and from the INAF main-stream funding programme (PI Vulcani). MLB acknowledges support from the National Science and Engineering Research Council 
(NSERC) Discovery Grant program.  GW gratefully acknowledges support from the National Science Foundation through grant AST-1517863, from HST program number GO-15294,  and from grant number 80NSSC17K0019 issued through the NASA Astrophysics Data Analysis Program (ADAP). Support for program number GO-15294 was provided by NASA through a grant from the Space Telescope Science Institute, which is operated by the Association of Universities for Research in Astronomy, Incorporated, under NASA contract NAS5-26555. 
We thank the International Space Science Institute (ISSI) for providing financial support and a meeting facility that inspired insightful discussions for team "COSWEB: The Cosmic Web and Galaxy Evolution". The Millennium Simulations databases used in this paper and the web application providing online access to them were constructed as part of the activities of the German Astrophysical Virtual Observatory. For this work we made use of the softwares Numpy \citep{numpy} and Scipy \citep{SciPy}.
\end{acknowledgements}

\bibliography{master}

\appendix
\section{CLUMPS}
\label{s:CLUMPS}
In this section we present a new method, called ~CLUster
Membership in Phase Space (CLUMPS), that is calibrated on numerical
simulations, and for which particular care is given to poor
systems. The main idea of this approach is that the exact
position of a galaxy in the phase space plane is not
fundamental. Rather, what matters is that many galaxies cluster around
the zero-velocity value of this plane.

For this reason the galaxy density field in the phase-space plane of
a cluster is computed in a grid, counting the number of galaxies that
lie in each bin of the grid. To construct this grid we need to
specify the step sizes $dR$ and $dV$ in the radial and velocity
directions, respectively. These values are converted into an integer
number of bins by rounding up the ratio of the extent to the
step in each direction. Specifically, the density field is
computed well outside the range of the data: in the radial direction
the grid starts at 0.999999 times the minimum radius up to twice the maximum
radius, while in the velocity direction from twice the minimum
velocity up to twice the maximum, doubling the number of bins in both
directions. This density field is then smoothed by convolving it with
a two-dimensional Gaussian kernel. To define the kernel we need to
specify the widths $\sigma_R$ and $\sigma_V$ of the Gaussian in the
radial and velocity directions, respectively.

Once the density field is smoothed, we consider the smoothed velocity
distribution in each radial bin. Since we assume that the velocity
center of the cluster is known exactly, the algorithm looks for the
maximum closest to zero and the two minima that enclose this peak. The
galaxies in each radial bin that have a velocity value within these
two minima are considered members. In all radial bins, except for the one closest to the cluster center, instead of requiring that the maximum must be the closest to zero,
it is required that the maximum must be the closest to the velocity
position of the maximum in the previous radial bin. This is done to
avoid jumps in the peak position from bin to bin due to the presence
of structures close to the zero-velocity. To find the minima we make use
of the SciPy \citep{SciPy} routine \texttt{argrelmin}, which has a
parameter to set how deep a minimum must be to actually be considered
as such. This is important for small fluctuations in the density not
to be considered  minima. We set this value to four. 

Here we note that the algorithm performance depends on four free
parameters, namely $dR$, $dV$, $\sigma_R$, and $\sigma_V$. They
are chosen by applying the algorithm with different values of such
parameters to a mock catalog, for which the true membership is
known. By comparing the latter with the one predicted by the algorithm it
is possible to choose the set of parameters that provides the
best results. In particular, we retrieved data from the Millennium database
\citep{Springel+05, Lemson+06} searching the \citet{Henriques+15}
light cones. We first downloaded a catalog of
halos satisfying the  conditions $\mvir > 10^{14} \rm
M_\odot/h$ and $0.1 < z < 0.5$. We then created fields of view (FOVs) that are centered
on halos with $1 < RA < 89$ deg and $-24 < Dec < 24$ deg, and
considered all the galaxies that satisfy the following condition: $0 <
RA < 90$ deg and $-25 < Dec < 25$ deg, $0.1 \le z_{\rm obs} \le 0.5$,
$z_{\rm obs}$ being the observed redshift, therefore including
peculiar velocities. The center of the FOVs is chosen in a smaller
area so that we do not consider FOVs too close to the border, avoiding the risk of a halo that is cut by the area edge. For the same reason, the FOV
center in redshift is restricted to $0.11 < z < 0.49$.  For each FOV
we retain all the galaxies that lie within a circle of radius 6
Mpc. To avoid FOVs with  too few objects, we retain FOVs
that have at least ten galaxies within 1 virial radius, and with
$|\vrf| < 6000$ km/s. This catalog contains 10094 FOVs.

To find the fiducial set of parameters, we run the algorithm on the
simulated sample varying the parameters. For each set of parameters
and for each FOV we consider all the galaxies with $|\vrf| < 4000$ km/s and within 3 Mpc from the
cluster center, and compute the completeness C and the purity P,
defined as 

\begin{equation}
  C_i = \frac{N_{\rm mem \cap <Rvir \cap CLUMPS}}{N_{\rm mem \cap <Rvir}}
,\end{equation}
\begin{equation}
  P_i = \frac{N_{\rm mem \cap <Rvir \cap CLUMPS}}{N_{\rm CLUMPS \cap <Rvir}}
,\end{equation}
where $N_{\rm mem \cap <Rvir \cap CLUMPS}$ is the number of members
within 1 virial radius that are also identified as members by CLUMPS,
$N_{\rm mem \cap <Rvir}$ is the number of members within 1 virial
radius, and $N_{\rm CLUMPS \cap <Rvir}$ is the number of galaxies
within 1 virial radius that are considered members by CLUMPS. We also
measure the velocity dispersion of the candidate members and that of
the actual members. Then, for each run, we divide the sample in bins
of richness (number of candidate members within 1 virial radius), and
for each bin we compute the median completeness, the median purity,
and the median value of the velocity dispersion divided by the true
velocity dispersion.

We find a good compromise among these quantities when using $dR =
0.075$ Mpc, $dV = 105$ km/s. For $\sigma_R$ and $\sigma_V$, instead
of using single values, we proceed as follows. For each run the code
is actually run 100 times, varying $\sigma_R$ from 1 to 6 in ten
linearly separated values, and $\sigma_V$ from 1 to 5 in ten linearly
separated values. The galaxies identified as members in at least half
of these runs are considered as final members.

With such a configuration we find that for the systems with
(estimated) richness between 30 and 50, the completeness is 0.95, the
purity 0.71, and the velocity dispersion ratio 0.97. When considering
richer systems (between 50 and 100) the completeness increases to
0.97, the purity decreases to 0.67, and the velocity dispersion ratio
is 0.96. 

CLUMPS therefore appears to perform reasonably well for systems with a number of members in the range of our 14 clusters, $28 \leq N_m \leq 86$ (see Table~\ref{t:data}). In particular, the fact that the cluster velocity dispersion is recovered with 5\% accuracy is particularly important for the dynamical analysis that relies on the velocity distribution of cluster galaxies.

\end{document}